\documentclass[acmsmall,screen,nonacm]{acmart}

\usepackage[utf8]{inputenc}
\usepackage[framemethod=TikZ]{mdframed}

\usepackage[utf8]{inputenc}
\usepackage{anyfontsize} 
\usepackage{semantic}    
\usepackage{stmaryrd}   
\usepackage{mathpartir}  
\usepackage{braket}      
\usepackage{thmtools}    
\usepackage{xspace}
\usepackage{proof}
\usepackage{myyhmath}
\usepackage{rotating}
\usepackage{tikz}
\usepackage{changepage}
\usepackage{listings}
\usepackage{mdframed}
\usepackage{mathtools}
\usepackage{ dsfont }
\usepackage[toc,page]{appendix}
\usepackage{pifont}
\usepackage[outline]{contour}
\usepackage{makecell}
\usepackage{stackengine,graphicx}
\usepackage{multicol}
\usepackage{stackrel}
\usepackage{extarrows}
\usepackage{tabularx}
\usepackage{utfsym}
\usepackage{mdframed} 
\usepackage[capitalise]{cleveref}
\usepackage{xcolor}
\usepackage{tikz}
\usetikzlibrary{matrix,positioning,arrows.meta}

\usetikzlibrary{shapes.geometric, arrows}

\tikzstyle{mynode} = [rectangle split, rectangle split parts=2, rounded corners, minimum width=3cm, minimum height=1cm,text centered, draw=black]
\tikzstyle{myarrow} = [->, >=stealth, double]
\tikzstyle{mytriangle} = [diamond, text centered, minimum width=3cm, minimum height=1.3cm, draw]

\crefname{section}{\textsection\!}{\textsection\!}
\crefname{figure}{Fig.}{Figs.}      
\Crefname{figure}{Figure}{Figures}  

\usetikzlibrary{automata, graphs,positioning,chains,arrows,decorations.pathmorphing}

\usetikzlibrary{arrows,positioning,backgrounds}

\newcommand{\lang}{\textsf{GTLC+}\xspace}%

\newcommand{\ie}{\emph{i.e.}\xspace}
\newcommand{\etal}{\emph{et al.}\xspace}
\newcommand{\eg}{\emph{e.g.}\xspace}

\usepackage{tabularx}

\definecolor{violeta}{HTML}{9000FF}
\definecolor{propcolor}{HTML}{3F7D31}
\definecolor{etcolor}{HTML}{ED720E}

\newcommand{\node}[1]{\llparenthesis#1\rrparenthesis}
\newcommand{\mynote}[3]
   {{\color{#3} \fbox{\bfseries\sffamily\scriptsize#1}
   {\small$\blacktriangleright$\textsf{\emph{#2}}$\blacktriangleleft$}}~}

\definecolor{mynotecolor}{HTML}{F282B4}

\definecolor{customgreen}{HTML}{006400}

\newcommand{\et}[1]{\mynote{ET}{#1}{blue}}
\newcommand{\mt}[1]{\mynote{MT}{#1}{red}}

\newif\ifdiffmode
\diffmodefalse

\newenvironment{changedenv}
  {\begingroup\color{blue}}  
  {\endgroup}

\newcommand{\evcolor}[1]{{\color{blue} #1}}
\newcommand{\icolor}[1]{{\color{blue} #1}}

\newcommand{\?}{\textsf{\upshape ?}}

\newcommand{\cT}{G} 
\newcommand{\cE}{E} 

\newcommand{\Int}{\mathsf{Int}}
\newcommand{\String}{\mathsf{String}}

\newcommand{\Bool}{\mathsf{Bool}}
\newcommand{\ev}[1][]{\evcolor{\varepsilon_{#1}}}
\newcommand{\pr}[1]{\evcolor{\braket{{#1}}}}
\newcommand{\interior}[1]{\mathcal{I}_{#1}}
\newcommand{\meet}{\sqcap}
\newcommand{\false}{\mathsf{false}}

\newcommand{\D}{\mathcal{D}}

\newcommand{\emptyenv}{\ensuremath{\o}\xspace}
\newcommand{\trans}[1]{\mathbin{\evcolor{\circ^{#1}}}}
\newcommand{\red}{\longmapsto}

\newcommand{\error}{\textup{\textbf{error}}}

\DeclareDocumentCommand{\itt}{ m O{} }{\icolor{\IfNoValueTF{#2}{t^{{#1}}}{t^{{#1}}_{#2}}}}
\DeclareDocumentCommand{\ittp}{ m O{} }{\icolor{\IfNoValueTF{#2}{t'^{{#1}}}{t'^{{#1}}_{#2}}}}
\newcommand{\TermT}[1]{\icolor{\mathds{T}}[#1]}
\newcommand{\ix}[1][]{\icolor{x^{{#1}}}}
\newcommand{\VarT}[1]{\mathds{V}[#1]}
\DeclareDocumentCommand{\itr}{ O{} }{\icolor{\IfNoValueTF{#1}{t}{t_{#1}}}}
\DeclareDocumentCommand{\itrp}{ O{} }{\icolor{\IfNoValueTF{#1}{t'}{t'_{#1}}}}
\newcommand{\iapp}[1]{\mathbin{\icolor{@^{#1}}}}
\newcommand{\braketeq}[2]{\evcolor{\braket{#1}}}
\newcommand{\iasc}[2]{\icolor{#1 :: #2}}

\newcommand{\relsymbol}{R}
\DeclareDocumentCommand{\rel}{m m}{\relsymbol(#1,#2)}
\DeclareDocumentCommand{\crel}{m m}{\widetilde{\relsymbol}(#1,#2)}
\newcommand{\dom}{\mathit{dom}}
\newcommand{\cod}{\mathit{cod}}

\newcommand{\boxit}[1]{\text{\lstinline{ref}}~{#1}}
\newcommand{\unbox}[1]{\text{\lstinline{!}}{#1}}
\newcommand{\assign}[2]{{#1} := {#2}}
\DeclareDocumentCommand{\vector}{m m}{\text{\lstinline{vec}}~{#1}~{#2}}
\DeclareDocumentCommand{\vectorset}{m m m}{\text{\lstinline{vset}}~{#1}~{#2}~{#3}}
\DeclareDocumentCommand{\vectorget}{m m}{\text{\lstinline{vget}}~{#1}~{#2}}
\DeclareDocumentCommand{\tuple}{m}{(#1)}
\DeclareDocumentCommand{\proj}{m m}{{#1}.(#2)}
\DeclareDocumentCommand{\letin}{m m m}{\text{\lstinline{let}}~{#1} = {#2}~\text{\lstinline{in}}~{#3}}
\DeclareDocumentCommand{\letrecin}{m m m}{\text{\lstinline{let rec}}~{#1} = {#2}~\textsf{in}~{#3}}
\DeclareDocumentCommand{\ifthenelse}{m m m}{\text{\lstinline{if}}~{#1}~\text{\lstinline{then}}~{#2}~\text{\lstinline{else}}~{#3}}
\DeclareDocumentCommand{\loopit}{m m m m}{\text{\lstinline{for}}~{#1}~=~{#2}~\text{\lstinline{to}}~{#3}~\text{\lstinline{do}}~{#4}~\text{\lstinline{done}}}
\DeclareDocumentCommand{\fun}{m m}{\text{\lstinline{fun}}~({#1}) -> {#2}}

\DeclareDocumentCommand{\binop}{m m}{#1~\odot~#2}
\DeclareDocumentCommand{\datastruct}{m}{\text{\lstinline{type}}~{#1}~=~\evcolor{[}|~ C~\text{\lstinline{of}}~G \evcolor{]^{+}}}
\DeclareDocumentCommand{\match}{m}{\text{\lstinline{match}}~{#1}~\text{\lstinline{with}}~\evcolor{[}|~C~x\ldots \text{\lstinline{->}}~e\evcolor{]^{+}}}

\DeclareDocumentCommand{\tvec}{m}{\text{\lstinline{vec}}[{#1}]}
\DeclareDocumentCommand{\tint}{}{\text{\lstinline{int}}}
\DeclareDocumentCommand{\tbool}{}{\text{\lstinline{bool}}}

\DeclareDocumentCommand{\tfloat}{}{\text{\lstinline{float}}}
\DeclareDocumentCommand{\tunit}{}{\text{\lstinline{unit}}}
\DeclareDocumentCommand{\tunk}{}{\text{\lstinline{?}}}
\DeclareDocumentCommand{\tfun}{m m}{{#1} \to {#2}}
\DeclareDocumentCommand{\tref}{m}{\text{\lstinline{ref}}[{#1}]}

\DeclareDocumentCommand{\boxedtwo}{m m}{
  {
  \setlength{\fboxsep}{0pt}  
  \fbox{%
  \begin{tabular}{c|c}
    #1 & #2
  \end{tabular}%
  }
  }
}
\DeclareDocumentCommand{\boxedthree}{m m m}{
  {
  \setlength{\fboxsep}{0pt}  
  \fbox{%
  \begin{tabular}{c|c|c}
    #1 & #2 & #3
  \end{tabular}%
  }
  }
}

\newcommand{\true}{\mathtt{true}}

\lstdefinestyle{ocaml}{
  language=[Objective]Caml,      
  basicstyle=\ttfamily\small,    
  numbers=left,                  
  numberstyle=\tiny,
  stepnumber=1,
  showstringspaces=false,
  columns=fullflexible,
  keepspaces=true,
  frame=single,
  framerule=0.3pt,
  breaklines=true,
  tabsize=2,
  keywordstyle={\bfseries\color{blue!70!black}},
  keywordstyle=[2]{\bfseries\color{teal!70!black}},
  commentstyle=\itshape\color{green!50!black},
  stringstyle=\color{purple!60!black},
  morekeywords={
    int,bool,float,char,string,unit,ref,mutable,fun,let rec,match,with,if,then,else,type,module,struct,open,include, vec, type, match, with, vget, vset, box, unbox,boxed, bflt, iflt, ifloat, bfloat, imm, box
  },
  morekeywords=[2]{dom,cod,ascribe,self},
  literate=
    {λ}{{$\lambda$}}1
    {->}{{$\to$}}2
    {=>}{{$\Rightarrow$}}2
    {>=}{{$\ge$}}2
    {<=}{{$\le$}}2
    {?}{{\bfseries\color{blue!70!black}?}}1
    {<}{{$\langle$}}1
    {>}{{$\rangle$}}1
}
\newcommand{\oblset}[1]{\textsc{#1}}

\newcommand{\Type}{\oblset{Type}}

\newcommand{\bench}[1]{\textsf{#1}}

\graphicspath{
  {benchs/final} 
  {benchs/2025-11-03-opt-prec-v3/} 
  {figures/} 
}

\title{Compiling Gradual Types with Evidence} 

\author{Jos\'e Luis Romero}
\affiliation{%
  \institution{PLEIAD Lab, Computer Science Department (DCC), University of Chile}
  \city{Santiago}
  \country{Chile}
}
\email{joromero@dcc.uchile.cl}

\author{Crist\'obal Isla}
\affiliation{%
  \institution{PLEIAD Lab, Computer Science Department (DCC), University of Chile}
  \city{Santiago} 
  \country{Chile}
}
\email{cisla@dcc.uchile.cl}

\author{Mat\'ias Toro}
\affiliation{%
  \institution{PLEIAD Lab, Computer Science Department (DCC), University of Chile}
  \city{Santiago}
  \country{Chile}
}
\orcid{0000-0002-5315-0198}
\email{mtoro@dcc.uchile.cl}

\author{\'Eric Tanter}
\affiliation{%
  \institution{PLEIAD Lab, Computer Science Department (DCC), University of Chile}
  \city{Santiago}
  \country{Chile}
}
\orcid{0000-0002-7359-890X}
\email{etanter@dcc.uchile.cl}

\begin{document}

\begin{abstract}
Efficiently supporting sound gradual typing in a language with structural types is challenging. To date, the Grift compiler is the only close-to-the-metal implementation of gradual typing in this setting, exploiting coercions for runtime checks, and further extended with monotonic references for efficient access to statically-typed data structures. On the language design and semantics side, the Abstracting Gradual Typing (AGT) methodology has proven fruitful to elucidate existing designs and to innovate by deriving gradualizations of a wide variety of typing disciplines and language features. Grounded in abstract interpretation, the Curry-Howard inspired runtime semantics of AGT is based on the notion of evidence for consistent judgments that evolve during reduction, monitoring the plausibility of well-typedness. While expressive and versatile, it is unclear whether such evidence-based semantics are a viable route to realize an efficient implementation of gradual typing.

In this work, we explore this question by designing, implementing, and evaluating an evidence-based compiler, called GrEv. We explain how to bridge the gap between the formal semantics and the GrEv compiler implementation, and identify novel monotonic semantics. We empirically evaluate the performance of GrEv on the Grift benchmark suite. The results show that an evidence-based compiler can be competitive with, and even faster than, a coercion-based compiler, exhibiting more stability across configurations on the static-to-dynamic spectrum. In addition to enriching the space of gradual typing compilers, this work opens a direct door to exploring efficient implementations of the many advanced gradual typing disciplines formally derived with AGT in the literature.

\end{abstract}

\maketitle

\section{Introduction}
\label{sec:intro}

Gradual typing reconciles static and dynamic typing by supporting imprecise types, statically checking as much as possible, treating imprecision optimistically, and deferring optimistic assumptions to runtime checking~\cite{siekTaha:sfp2006,siekAl:snapl2015}. The literature has explored a wide range of formal and implementation approaches to gradual typing, either by retrofitting a gradual type system on top of an existing dynamically-typed language~\cite{tobinFelleisen:popl2008,vitousekAl:dls2014,vitousekAl:popl2017}---a trend that is growing in industry with languages such as TypeScript among others---or by relaxing a statically-typed discipline with imprecision, as originally conceived~\cite{siekTaha:sfp2006}. \citet{takikawaAl:popl2016} warned of the potentially enormous overheads incurred by gradual typing in certain configurations, fostering further research into gradual typing variations and implementation techniques~\cite{kuhlenschmidtAl:pldi2019,toroTanter:scp2020,tsudaA:ecoop2020,vitousekAl:popl2017,vitousekAl:dls2019,muehlboeckTate:oopsla2017,muehlboeckAndTate:oopsla21,camporaAl:popl2024}. 
Most notable dimensions are whether the type system is structural~\cite{siekTaha:sfp2006,siekTaha:ecoop2007,garciaAl:popl2016} or nominal~\cite{muehlboeckTate:oopsla2017,muehlboeckAndTate:oopsla21}, whether gradual typing is coarse grained (per module)~\cite{tobinFelleisen:popl2008} or fine grained (per type annotation), whether the soundness guarantees are deep, shallow or optional~\cite{greenmanFelleisen:icfp2018}, whether the language is functional or object-oriented~\cite{siekTaha:ecoop2007,takikawaAl:oopsla2012}, and whether gradual typing is built on top of a dynamically-typed or statically-typed backend---in the latter case, simply ignoring type information as is done in TypeScript is not an option if one expects a safe language as a result.

In this work, we focus on structural, fine-grained sound gradual typing for a higher-order language with mutable data structures, in which statically-typed code can be compiled efficiently, without runtime type checks. This setting is important to study if and how one could achieve an efficient implementation of a gradually-typed version of OCaml, for instance. While achieving efficient integration in a production-quality compiler is still an open issue, there have been several advances in recent years. 
Most notably, Grift~\cite{kuhlenschmidtAl:pldi2019} has become the reference point to study close-to-the-metal implementations of gradual typing and their optimizations. Grift shows the benefits of using coercions~\cite{hermanAl:hosc10} over type-based casts. Subsequently, \citet{almahallawi:phd} extends Grift to evaluate the performance of monotonic references~\cite{siekAl:esop2015}, a discipline that ensures that runtime type information of heap-allocated structures can only become more precise over time, making it possible to further optimize manipulation of statically-typed data, \citet{tsudaA:ecoop2020} explore coercion-passing style in Grift for recovering tail call optimization, and recently, \citet{camporaAl:popl2024} study an optimization technique called discriminative typing by implementing it in Grift.

In parallel with work on implementation techniques, the gradual typing literature is also rich in language design and semantics exploration of gradual typing for different language features (mutable references~\cite{siekTaha:sfp2006,siekAl:esop2015}, objects~\cite{siekTaha:ecoop2007,takikawaAl:oopsla2012,allendeAl:scp2014}, polymorphism~\cite{igarashiAl:icfp2017, ahmedAl:icfp2017,toroAl:popl2019,ina:oopsla2011}, delimited continuations~\cite{miyazakiAl:stop2016}, effect handlers~\cite{newAl:oopsla2023}, etc.) and for advanced typing disciplines (effect typing~\cite{banadosAl:icfp2014}, typestates~\cite{wolffAl:ecoop2011}, security typing~\cite{thiemannFennell:esop2014}, ownership~\cite{sergeyClarke:esop2012}, refinement types~\cite{lehmannTanter:popl2017}, etc.).
In that space, the Abstracting Gradual Typing (AGT) methodology~\cite{garciaAl:popl2016} has proven effective 
to ground gradual language design in firm principles: based on abstract interpretation~\cite{cousot:popl1977}, most specifically in viewing gradual types as abstractions of sets of static types, AGT can drive the gradualization of an existing static typing discipline, by lifting static type relations and operators to their consistent counterpart through a Galois connection.  AGT gives a Curry-Howard account of runtime checking as the consequence of evolving {\em evidence} of consistent judgments in typing derivations as a program reduces via a {\em consistent transitivity} operator, in spirit replaying the type safety proof of the language at runtime. 

AGT has been used both to elucidate existing designs and to drive innovation in gradual typing. To this date, notions of evidence and consistent transitivity for a wide variety of features and typing disciplines have been studied~\cite{banadosAl:jfp2016,lehmannTanter:popl2017,toroTanter:sas2017,toroAl:toplas2018,toroTanter:scp2020,toroAl:popl2019,banados:popl2021,labradaAl:oopsla2022,yeAl:oopsla2023,arquezAl:csf2025}.
While solid and expressive, this approach is not designed for efficiency, but rather derived from the metatheory.
Formal presentations of evidence-based semantics for AGT languages have evolved from a reduction of typing derivations, or annotated intrinsically-typed terms, towards a simpler, closer to an implementation presentation 
in which all values are decorated with evidence, consistent judgments are restrained to ascriptions, and the runtime semantics systematically apply consistent transitivity to combine ascriptions~\cite{toroTanter:scp2020,toroAl:popl2019}.
This simpler presentation, also adopted in more recent work~\cite{arquezAl:csf2025,yeAl:oopsla2023}, suggests a rather direct implementation strategy, albeit potentially naive and inefficient. Given the expressive advantages of the AGT conceptual framework, we wonder whether a direct implementation of an evidence-based semantics in a compiler would indeed be utterly inefficient, or if it could be competitive with coercions. In this work, we settle to answer this question by building a new compiler for a gradually-typed language with first-class functions and mutable data structures, based on evidence. 

Specifically, this article makes the following contributions:
\begin{itemize}
\item We explain how to bridge the gap between formalism and low-level implementation of evidence-based semantics, in particular the low-level representation of evidence values, and the realization of evidence-based reduction rules in a compiler. We explain how some basic gradual typing optimizations from Grift can be realized in the evidence-based setting.
\item We describe different designs corresponding to the \emph{guarded} and \emph{monotonic} semantics from the literature~\cite{siekAl:esop2015}, and identify a novel monotonic semantics---\emph{monotonic values}---in which all heap-allocated values, in particular closures, are treated monotonically.
\item We implement a new ahead-of-time compiler, called GrEv, which targets LLVM IR and shares the same source language as Grift. GrEv is implemented in OCaml, with some runtime components written in C.  
GrEv supports three semantics: guarded values, monotonic data structures (with guarded closures), and monotonic values.  
\item We empirically evaluate the performance of GrEv compared to Grift across all guarded and monotonic semantic variants, on the Grift benchmark suite~\cite{kuhlenschmidtAl:pldi2019,almahallawi:phd}.
The results show that an evidence-based compiler can achieve performance competitive with, and even superior to, a coercion-based compiler, and appears more stable across configurations on the static-to-dynamic spectrum. The comparison of the different semantic modes of both compilers confirms the interest of monotonic approaches to gradual typing. We also report on an optimization for floats, present in Grift and optional in GrEv, whose impact is observedly contentious. 
\end{itemize}
Finally, while this article is only concerned with gradualizing a simple type system, an evidence-based compiler such as GrEv opens the door to exploring the efficient implementation of advanced gradual typing disciplines explored in the AGT literature~\cite{banadosAl:jfp2016,lehmannTanter:popl2017,toroTanter:sas2017,toroAl:toplas2018,toroTanter:scp2020,toroAl:popl2019,banados:popl2021,labradaAl:oopsla2022,yeAl:oopsla2023,arquezAl:csf2025}.

\cref{sec:background} provides background on gradual typing and Abstracting Gradual Typing (AGT), \cref{sec:grev} describes the design and implementation of the evidence-based compiler GrEv with its semantic variations, and \cref{sec:performance} reports on the performance evaluation of GrEv and Grift. We finally discuss related work (\cref{sec:related}) and conclude (\cref{sec:conclusion}).
The implementation of GrEv, as well as the experimental artifacts used in the performance evaluation, are available online at \url{https://pleiad.cl/grev/}.






\section{Background on (Abstracting) Gradual Typing}
\label{sec:background}

We start with a brief introduction to gradual typing (\cref{sec:bg-gt}), followed by the main ingredients of the Abstracting Gradual Typing methodology (\cref{sec:bg-agt}), specifically evidence-based semantics (\cref{sec:ev-semantics}).

\subsection{Gradual Typing}
\label{sec:bg-gt}

Static and dynamic typing provide complementary benefits and drawbacks. Static typing detects errors early but may conservatively reject valid programs, while dynamic typing offers flexibility at the cost of runtime checks and potential errors.
Gradual typing combines both disciplines in a single language, allowing programmers to locally choose between static and dynamic checking and to evolve programs smoothly along this spectrum~\cite{siekTaha:ecoop2007}.

Gradual typing formalizes this flexibility through a key relation between gradual types called \emph{(im)precision} ($\cT_1 \sqsubseteq \cT_2$), which  intuitively captures the amount of static information provided by a type. The most imprecise type is usually called the unknown type $\?$, while a static type is a most precise type. Structural gradual typing allows occurrences of the unknown type at any level of the structure of a type, \eg~$\? \to \Bool$ to denote functions that definitely return a boolean value, but accepts values of any type as argument.
A gradual type system treats imprecision optimistically, by using type relations that are tolerant to imprecision. For instance, type {\em consistency} ($\sim$) relaxes standard equality by only requiring that two types agree on their known parts, \eg~$\? -> \Bool \sim \Int -> \?$.
Importantly, consistent relations are generally not transitive: $\Int \sim \?$ and $\? \sim \Bool$, but $\Int \not\sim \Bool$, otherwise no program would be statically deemed ill-typed. For more advanced type systems, \eg~with subtyping, one needs to devise optimistic counterparts, such as consistent subtyping~\cite{siekTaha:ecoop2007}. 

Consider the following example that illustrate three different scenarios in gradual typing:
\begin{lstlisting}
let f : ?->int = fun g -> g(1) in                            (* add1: int -> int *)
f(add1)        (* typechecks and runs without errors *)
not(f(add1))   (* ill-typed *) 
f(true)        (* typechecks and fails at runtime *)
\end{lstlisting}
First of all, the body of \lstinline|f| typechecks because \lstinline|g| has type $\tunk$, which is consistent with the most permissive function type $\tunk->\tunk$, and $\tunk$ itself is consistent with the declared return type, $\tint$.
On line 2, the application \lstinline|f(add1)| typechecks 
because $\tint->\tint \sim \tunk$, and successfully evalutes to 2. On line 3, the application of \lstinline|not|, which expects a boolean, is statically rejected because \lstinline|f| is statically known to return an $\tint$. Finally, the application on line 4 is well-typed because $\tbool \sim \tunk$, but ought to fail at runtime prior to applying \lstinline|true| as a function.


Ensuring that line 4 fails dynamically is at the crux of the challenges of realizing gradual typing in practice. If one is implementing gradual typing as a retrofitted type system on top of a safe dynamic language (\eg~TypeScript or Typed Racket), then one could be satisfied with letting the underlying runtime system ensure safety; an approach known as optional typing~\cite{bracha:rdl2004}. If, however, one is adding gradual types to a statically-typed language like OCaml, then the gradual typing implementation must take responsibility for ensuring safety, performing a runtime check before optimistically accepted operations, like the application \lstinline|g(1)| above.

The seminal presentation of gradual typing~\cite{siekTaha:sfp2006} follows the latter route, and devises the semantics of the gradual language via elaboration to an internal language 
with explicit runtime tagging and checking, known as a \emph{cast calculus}. Casts are inserted at boundaries between statically known and unknown information, ensuring that any violation of static assumptions is detected and reported as a runtime error, instead of producing segmentation faults. 
A possible elaboration of the program above, considering the use on line 4, would be (\lstinline|e : G => G'| denotes the cast of \lstinline|e| from \lstinline|G| to \lstinline|G'|):
\begin{lstlisting}[numbers=none]
let f : ?->int = fun g -> (g : ? => ?->?)(1 : int => ?) : ? => int in 
f(true : bool => ?)
\end{lstlisting}
Note how casts are inserted at all places where consistency is used to deem the program gradually well typed. Upcasts to the unknown type can be understood as ``tagging'' operations, while downcasts from the unknown type to a more precise type constitute runtime checks, which can read previously tagged information. Here during reduction, after substituting \mbox{\lstinline{true : bool => ?}} (a tagged boolean) in the body of the function, the redex \mbox{\lstinline{false : bool => ? : ? => ?->?}} triggers a runtime error, since a boolean cannot be casted to a function type.

While the core principles of gradual typing exposed here—imprecision, consistency, and runtime checks—are rather straightforward, as mentioned in the introduction, the introduction of runtime checks can have dramatic impact on performance. In addition, scaling these principles to more complex languages, with more complex typing relations (such as subtyping, polymorphism, effects, dependencies, etc.) can become quite subtle to get right. Gradual typing presents serious challenges, both semantically and implementation-wise.

\subsection{Abstracting Gradual Typing}
\label{sec:bg-agt}
The Abstracting Gradual Typing (AGT) methodology provides a principled and general framework for systematically designing a gradually-typed language, based on abstract interpretation~\cite{cousot:popl1977}, starting from a statically-typed language. The key idea is to view gradual types as {\em abstractions} of sets of static types, formally defined through a Galois connection (with concretization $\gamma$ and sound and optimal abstraction $\alpha$). For instance, viewing the unknown type as an abstraction of any type ($\gamma(\?)=\Type$), and propagating this through all type constructors gives the standard interpretation of GTLC~\cite{siekTaha:sfp2006,siekAl:snapl2015},
with for instance~$\gamma(\?->\Int) = \{ T -> \Int \,|\, T \in \Type \}$ and $\alpha(\{\Int->\Bool, \Int->\String\}) = \Int ->
\?$. The notion of {\em precision} between gradual types~\cite{siekAl:snapl2015} follows naturally, as $G_1 \sqsubseteq G_2 \triangleq \gamma(G_1) \subseteq \gamma(G_2)$.

Armed with this Galois connection, which syntactically captures the meaning of gradual types, AGT drives the gradualization of a static typing discipline in two steps. First, for the static semantics, one simply lifts through the Galois connection all the type relations (\eg~equality, subtyping, containment, etc.) and partial type operators (\eg~equating, subtyping join, union, etc.). This lifting is existential, capturing the fact that gradual typing is about the {\em plausibility} of well-typedness. For instance, consistency is naturally understood as the existential lifting of equality: $G_1 \sim G_2 \triangleq \exists T_1 \in \gamma(G_1), T_2 \in \gamma(G_2), T_1 = T_2$,  meaning that both types denote at least one common static type.

Second, for the dynamic semantics, AGT exploits the Curry-Howard correspondence between proof normalization (of the type safety argument) and term reduction, in order to account for possible runtime errors: instead of just being concerned with plausibility, one tracks the local {\em evidence} of such plausibility, and reduction enforces these local justifications to be combined in order to attempt to justify transitive steps (a partial operation called {\em consistent transitivity}, also direct to define in terms of the Galois connection~\cite{garciaAl:popl2016}); incompatibility denotes a justified runtime type error. Given its central role in this work, we provide more details on evidence-based semantics below (\cref{sec:ev-semantics}). Here again, AGT has proven effective both at elucidating prior notions, such as threesomes~\cite{siekAl:popl10}, as well as devising new runtime semantic operators in challenging settings. 

Overall, AGT has been successfully used across a wide range of features and typing disciplines, including 
records and subtyping~\cite{garciaAl:popl2016,banados:popl2021}, type-and-effects~\cite{banadosAl:icfp2014,banadosAl:jfp2016}, refinement types~\cite{lehmannTanter:popl2017,vazouAl:oopsla2018}, set-theoretic and union
types~\cite{castagnaLanvin:icfp2017,toroTanter:sas2017}, information-flow typing~\citep{toroAl:toplas2018}, parametric polymorphism~\citep{toroAl:popl2019,labradaAl:oopsla2022,labradaAl:jacm2022}, flexible data types~\cite{malewskiAl:oopsla2021}, probabilistic lambda calculus~\citep{yeAl:oopsla2023}, and sensitivity typing~\cite{arquezAl:csf2025}.














\subsection{Evidence-Based Semantics}\label{sec:ev-semantics}
As mentioned above, instead of giving the dynamic semantics of a gradual language by translation to an cast calculus~\cite{siekTaha:sfp2006}, AGT provides a {\em direct} dynamic semantics of gradual programs, defined over gradual typing derivations~\cite{garciaAl:popl2016}. The key idea is to apply {\em proof reduction} on gradual typing derivations~\cite{howard80formulae} augmented with {\em evidence} for consistent judgments, which---by the Curry-Howard correspondence---induces a canonical notion of reduction for gradual terms. 

To get an intuition of how this works, consider the term $t(v) := (\lambda x:\?. x+1)\; v$, which is gradually well-typed but may fail at runtime depending on what $v$ is, and its typing derivation $\D$:

\begin{small}
\begin{mathpar}
\inference{
     \inference{
      \inference{
        \inference{x: \? \in x: \?}{x: \? |- x : \?} &
        \inference{}{x: \? |- 1 : \Int} &
        \ev[1] |- \? \sim \Int 
      }{x: \? |- x + 1: \Int }
    }{\emptyenv |- (\lambda x:\?. x+1): \? -> \Int}
     & 
    \inference{}{\emptyenv |- v : G_v} &
    \ev[2] |- G_v \sim \?
  }{\emptyenv |- (\lambda x:\?. x+1)\; v : \Int}  
\end{mathpar}
\end{small}

\noindent In the derivation, we have left opaque the {\em evidences} of the two consistency judgments that allow this program to be gradually well typed.
Logically, a proof that uses an implication can be normalized by substituting the proof of the assumption for the use of the variable in the implication proof---the Curry-Howard equivalent of $\beta$-reduction. The type preservation argument for this step, in the static setting, (implicitly) exploits the transitivity of equality but here, consistency has been used instead, and transitivity does not hold universally for consistency. Therefore, the reduction step requires establishing a new consistency judgment via a transitivity step between the consistent judgments of $\D$; whether or not this is legal depends on the local justifications of both judgments, namely $\ev[1]$ and $\ev[2]$: if they can be combined by {\em consistent transitivity}, noted $\ev[2] \circ \ev[1]$, then the reduction is allowed to proceed, using the produced evidence as justification in the resulting derivation tree:

\begin{small}
\begin{mathpar}
\D \red
\inference{
  \inference{}{\emptyenv |- v : G_v} &
  \inference{}{|- 1 : \Int} &
  \ev[2] \circ \ev[1] |- G_v \sim \Int
}{v + 1 : \Int}
\qquad \text{if $\ev[2] \circ \ev[1]$ defined}
\end{mathpar}
\end{small}

\noindent If $v:=10$, $G_v:=\Int$, all is fine. But if $v:=\false$, $G_v:=\Bool$, then $\ev[2] \circ \ev[1]$ is undefined, and therefore the reduction yields an error, \ie $\D \red \error$ if $\ev[2] \circ \ev[1]$ is undefined.

This principled presentation of the runtime semantics of gradual typing scales to richer languages and complex typing disciplines, as witnessed by the many formal systems constructed in this manner. Given an appropriate representation for evidence, AGT yields the definition of consistent transitivity. Then the language designer can come up with algorithmic definitions and prove them equivalent to the ``ground truth'' derived via the abstract interpretation framework. 

In this work, we are concerned with simple types, where the only consistent judgment is consistency, the existential lifting of type equality (which is symmetric). In this extremely simple setting, evidence can be represented as a single gradual type, abstracting the common denoted types in a consistency judgment. For instance, $\pr{\Int->\Bool} |- \? -> \Bool \sim \Int -> \?$. Likewise, consistent transitivity is just the precision meet between both types: $\pr{G_1} \trans{} \pr{G_2} = \pr{G_1 \meet G_2}$.

\paragraph{Simplified presentation.} 
To formalize the runtime semantics of gradual programs while avoiding writing reduction rules on general derivation trees, Garcia~\etal adopt {\em intrinsic terms}~\cite{church:jsl1940}, a flat notation isomorphic to typing derivations. While adequate, the resulting description
is fairly exotic and cumbersome to work with. Furthermore, it does implicitly involve a (presentational) transformation from source terms to their intrinsic representation. 

Recent work on AGT~\cite{toroAl:popl2019,labradaAl:jacm2022,arquezAl:csf2025,yeAl:oopsla2023} simplifies the exposition by avoiding the use of intrinsic terms. Instead, a simple type-directed elaboration first produces terms with explicit type ascriptions everywhere consistency is used---very much in the spirit of a coercion-based semantics of subtyping~\cite{pierce:tapl}. All values are ascribed with evidence, and only ascription expressions carry evidence. As a result, the formalized language is familiar, with typing rules and reduction rules that are greatly simplified. In particular, the typing rules of the judgment $\Gamma \vdash t : \cT$ are standard, except for the ascription rule, which uses the consistency judgment and carries its evidence:
\begin{mathpar}
  \inference[(eGapp)]{
          \Gamma |- t_1 : \cT_{11} -> \cT_{12} &
          \Gamma |- t_2 : \cT_{11}
      }{
      \Gamma |- t_1 ~ t_2 : \cT_{12}
      }
      \and
  \inference[(eG$::$)]{
          \Gamma |- t : \cT_1 &
          \ev |- \cT_1 \sim \cT_2 \\
      }{
      \Gamma |- \ev t :: \cT_2 : \cT_2
      }
\end{mathpar}


For evaluation, the target type of an ascription is computationally irrelevant, so values have the form $v::= \ev u$, and the reduction rule for function application and ascription are: 
\begin{align*}
      \ev[1] (\lambda x. t)~\ev[2] u 
      \red& \evcolor{\cod(\ev[1])} t[\evcolor{(\ev[2] \trans{} \dom(\ev[1]))}u/x] & \text{if defined, otherwise }
        \error         
      \\
      \ev[2] (\ev[1] u)
      \red&
      \evcolor{(\ev[1] \trans{} \ev[2])} u & \text{if defined, otherwise } \error
\end{align*}

To illustrate, consider the previous term $t(\false)$, its elaboration ($\leadsto$) and reduction:
\begin{small}
\begin{align*}
  t &\leadsto \pr{\? -> \Int}(\lambda x:\?. \pr{\Int} x + \pr{\Int} 1) ~ \pr{\Bool} \false\\
  &\red 
  \braketeq{\Int}{}(\braketeq{\Int}{} \braketeq{\Bool}{} \false + \braketeq{\Int}{}1) & (\evcolor{dom(\pr{\? -> \Int})}=\braketeq{\?}{} \text{ and } \braketeq{\Bool}{} \trans{} \braketeq{\?}{} = \braketeq{\Bool}{}) \\
  & \red \error  & (\braketeq{\Bool}{} \trans{} \braketeq{\Int}{} = \pr{\Bool \meet \Int}~\text{is undefined})
\end{align*}
\end{small}

This presentation suggests a rather direct strategy for implementing gradual typing with evidence, at least for proof-of-concepts interpreter.
Designing a compiler involves quite a few additional considerations, as explored in the rest of this article.

Finally, it is worth highlighting that, while here we work concretely on the specific case of simple gradual types with consistency (hence the simplified formal presentation above), the general framework and definitions of AGT and evidence-based semantics applies to a wide variety of typing disciplines. Developing a compiler based on evidence, where evidence representation and consistent transitivity implementations are treated abstractly by the rest of the compilation pipeline, constitutes an appealing general basis for studying the low-level implementations of a wide range of gradual typing disciplines. This work first addresses the question, in the simply-typed setting, of whether using evidence for compiling gradual programs can be competitive with coercions.

\section{The GrEv Compiler}
\label{sec:grev}

We explore the efficient realization of evidence-based gradual languages via the implementation of a compiler, named GrEv. 
Like Grift, GrEv focuses on structural, fine-grained, sound gradual typing, and targets a low-level, unsafe backend, meaning that the gradual typing machinery is responsible for performing type-related safety checks.

We start by briefly presenting the source language \lang and recalling basic elements of the compilation of such a language, considering both the statically-typed and dynamically-typed settings (\cref{sec:source}).
We then focus on the compilation of evidence-based gradual typing: evidence values (\cref{sec:ev-values}), elaboration and evaluation (\cref{sec:elab-eval}). This presentation of GrEv ends with a brief explanation of basic gradual typing optimizations (\cref{sec:optims}) and a comparison with Grift (\cref{sec:grev-grift}).

\subsection{Source Language and Basic Elements of Compilation}
\label{sec:source}

  \lang~\cite{kuhlenschmidtAl:pldi2019} is a core language with first-class functions, primitive types including integers, booleans, and floats, and different kinds of data structures: references (\ie~heap cells), vectors, immutable tuples, and variants (\ie~algebraic datatypes). Expressions include let-bindings (possibly recursive), application of primitive operators (such as addition, printing, etc.), function application, conditionals, loops, access to data structures, and type ascriptions.
  The syntax includes type annotations, which can be omitted. Types are structural, with type constructors for functions and data structures, as well as base types, and also include the unknown type $\?$. Recursive types in GrEv are iso-recursive, and only used for defining variants, 
  \eg~\lstinline|type stream = SCons of int * (int -> stream)|.%

  \begin{figure}[t]
    \begin{displaymath}
      \begin{array}{r@{\hspace{0.3em}}c@{\hspace{0.8em}}l@{\hspace{0.8em}}l}
        e & ::= & \fun{\overline{x: \cT}}{e} \mid e~e\ldots \mid x  & \text{(functions and variables)} \\
        & &  \mid n \mid f \mid \binop{e}{e} \mid \mathit{op}~e\ldots & \text{(integers, floats, and operators)} \\
        & &  \mid b \mid \ifthenelse{e}{e}{e}  & \text{(booleans and conditionals)} \\
        & &  \mid \loopit{x}{e}{e}{e} & \text{(loops)}\\
        & &  \mid\letin{x: \cT}{e}{e} \mid \letrecin{f: \cT}{e}{e} & \text{(let bindings)} \\
        & &  \mid\boxit{e} \mid \unbox{e} \mid \assign{e}{e} & \text{(references, unit)}\\
        &  &   \mid\vector{e}{e} \mid \vectorset{e}{e}{e} \mid \vectorget{e}{e}& \text{(vectors)} \\
        &  &  \mid\tuple{\overline{e}} \mid \proj{e}{e} & \text{(tuples)} \\
        & &  \mid \datastruct{N} \mid C~e\ldots  & \text{(variant definition and construction)} \\
        & &  \mid \match{e} & \text{(variant elimination)} \\
        & & \mid e :: \cT \mid e ; e& \text{(ascriptions and sequences)} \\
        \cT & ::= & \tint \mid \tbool \mid \tfloat \mid \tref{\cT} \mid \tvec{\cT} 
        & \text{(types)} \\ 
        & & \mid \tunit \mid \cT * \cT \ldots \mid \tfun{(\overline{\cT})}{\cT} \mid N \mid \tunk  &  
      \end{array}
    \end{displaymath}
    \caption{\lang source syntax}
    \label{fig:lang-syntax}
  \end{figure} 
  In addition to GrEv, which treats missing type annotations as $\tunk$ and supports sound gradual typing, we have implemented StaticGrEv, a standard compiler for \lang that demands all type annotations to be present and fully precise (no use of the unknown type), and follows the basic principles of compilation for a statically-typed language, as well as a fully dynamically-typed version, DynGrEv, which ignores all type annotations and proceeds as a standard dynamically-typed language compiler.
  In the statically-typed compiler, values of primitive types are immediate (meaning they can live on the stack and in registers), while data structures are heap-allocated and accessed via pointers. First-class functions are implemented as flat closures, \ie heap-allocated structures with the code pointer to the function body, and the values of lexically-captured variables (the environment). This compiler exploits the static type information to avoid any runtime type check.
  In contrast, in the dynamically-typed compiler, all accesses to values require tag checks for safety, so all values---whether immediate or heap allocated---are additionally marked with a type tag. This means that on a 64-bit architecture, integers are limited to $63$ bits, with one bit used as tag. Because floats cannot be reduced to $63$ bits, they have to be heap-allocated to be tagged, hence degrading performance of floating-point arithmetic compared to the statically-typed compiler which can treat floats as immediate values.

  The compilation pipeline of StaticGrEv is standard: lexing, parsing, typechecking, $\alpha$-renaming, ANF conversion, closure conversion (with some optimizations \cite{keepAl:sfp2012}), and lowering to LLVM IR. 

\subsection{Compiling with Evidence Values}
\label{sec:ev-values}

To implement a compiler for \lang based on evidence, we need to map the evidence-based formalism from \cref{sec:ev-semantics} to a low-level implementation strategy. 

The main ingredients are that all values are {\em evidence values}, $\ev u$, and that the elimination forms of the language (function applications, ascriptions, projections, etc.) need to combine the evidences of the involved values using {\em consistent transitivity}, $\ev[1] \trans{} \ev[2]$, in order to eventually produce an evidence value as a result, or raise an error if a runtime type inconsistency is detected.

\paragraph{Representing values} In the formalism, values are uniformly described, but in the implementation, some values are immediate (\eg integers)---meaning they can live on the stack or in registers---while others are heap-allocated (\eg~tuples and vectors). In a formal presentation, a heap is usually introduced only when one needs to account for mutation, but in a compiler, the fact that registers can only hold 64-bit words forces the use of the heap for any value that does not fit in one word.\footnote{Unless the value is local, in which case it could be allocated on the stack; both Grift and GrEv are simple compilers that do not attempt to allocate structures on the stack, for instance via an escape analysis.}

Furthermore, formal semantics usually realize first-class functions with lexical scoping $\lambda x. t$ via a substitution metafunction $t[v/x]$ that replaces each free occurrence of the formal parameter $x$ with the actual argument $v$ in the body $t$.
In contrast, any reasonably efficient implementation does not perform such a code transformation at runtime, but instead uses closures that capture their lexical environment. A closure cannot be implemented as an immediate value, because it requires at least a label (code pointer) and possibly several values for its captured lexical environment.

\paragraph{Evidence values} Because evidence for \lang can be a type of any size, one cannot in general represent evidence as a bit tag, and so a general uniform solution would be to place all evidence values on the heap, with their two components $\ev$ and $u$.
The main exceptions to this rule are integers and booleans. These evidence values can be realized as immediate values, because their associated evidence is necessarily their (atomic) type: \ie~if $v = \ev[]u$ and $u$ is an integer, then $\ev=\pr{\tint}$, independently of whether $v : \Int$ or $v : \?$. So one can implement the evidence of integers and booleans using bit tags, as usual. As mentioned in \cref{sec:source}, floats are more challenging because they cannot be both tagged and immediate; therefore, as a first approximation, in a gradual language floats need to boxed (we discuss a related optimization in \cref{sec:optims}).


\begin{table}[t]
\centering
\begin{footnotesize}
\begin{tabular}{@{}lll@{}}
\toprule
\textbf{Value} & \textbf{Monotonic} & \textbf{Guarded} \\
upon more precise
ascription: & update ev in place & shallow copy with new ev \\
\midrule
ref & \boxedtwo{ev}{val} & \boxedtwo{ev}{val\_ptr} \\[0.3em]
vector & \boxedthree{ev}{len}{vals...}  & \boxedthree{ev}{len}{vals\_ptr} \\[0.3em]
tuple & \boxedtwo{ev}{vals...} & \boxedtwo{ev}{vals\_ptr} \\[0.3em] 
variant & \boxedthree{ev}{ctr\_id}{vals...} & \boxedthree{ev}{ctr\_id}{vals...} \\[0.3em]
closure & \boxedthree{ev}{fun\_ptr}{vals...} & \boxedthree{ev}{fun\_ptr}{vals\_ptr} \\[0.3em]
\bottomrule
\end{tabular}
\end{footnotesize}
\caption{Representation of boxed evidence values in memory, in monotonic and guarded semantics. 
}
\label{tab:values}
\end{table}

\paragraph{Structures} 
Like in Grift, all data structures in GrEv are heap allocated. Let us focus on reference cells; other data structures are handled similarly. Conceptually, we can either put the evidence on the heap together with the cell data, or associate the evidence with the pointer to the cell. However, packing a pointer and an evidence does not fit in one word, so the latter actually implies using a (heap-allocated) proxy. These representation differences have semantic impact: the former corresponds to {\em monotonic references}, first proposed by \citet{siekAl:esop2015}, while the latter corresponds to the traditional {\em guarded references}~\cite{siekTaha:sfp2006}.


\begin{figure}[t]
  \hspace{1em}
\begin{minipage}[c]{0.26\linewidth}
\begin{lstlisting}[basicstyle=\ttfamily\footnotesize]
let x = ref (4 :: ?) in
let y: ref[bool] = x in
y := true; 
!y
\end{lstlisting} 
    
  \end{minipage}
  \hfill
  \begin{minipage}[c]{0.68\linewidth}
    \includegraphics[width=\linewidth]{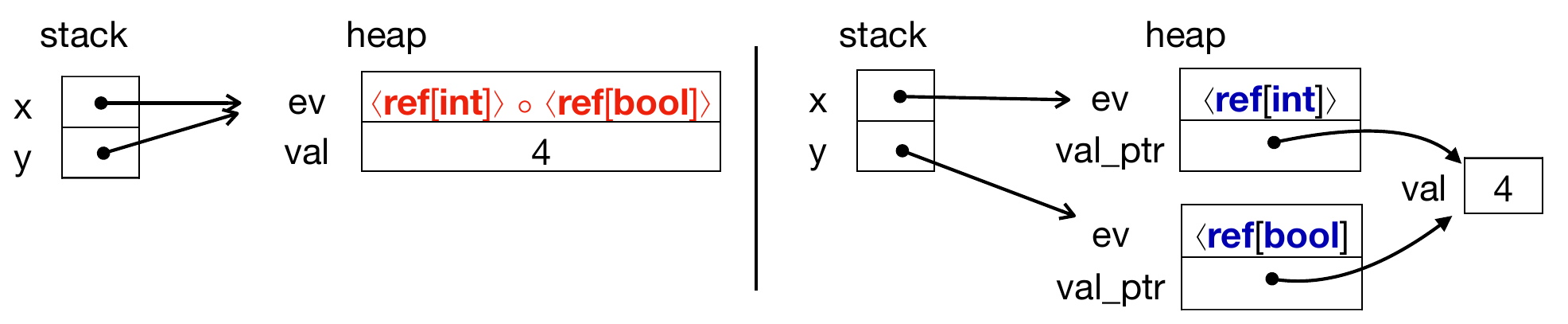}
  \end{minipage}
\caption{Monotonic vs. Guarded References}
\label{fig:mono-vs-guarded}
\end{figure}
With monotonic references, the evidence of a value in the heap is at least as precise as any reference that points to it. When an alias at a more precise type is introduced, the evidence on the heap is correspondingly updated, or an error is raised if an inconsistency is detected. This approach has two performance-related benefits: {\em (a)} because the runtime type information (evidence) is uniquely shared across all aliases, there is no need to create proxies, and {\em (b)} accesses to a statically-typed reference can dispense from runtime checks.
To illustrate the difference between monotonic and guarded references, consider the program shown in Figure~\ref{fig:mono-vs-guarded}.  When the reference bound to $x$ is created, with a value of unknown static type, its associated evidence in the heap is $\pr{\tref{\tint}}$.  
Later, when $y$ is created with type $\tref{\tbool}$, this information is combined with the current evidence of the heap cell, which here leads to a runtime error at line~2.

In contrast, guarded references are more flexible for the programmer but require every access to be dynamically checked, even from statically-typed code.
When $x$ and $y$ are introduced, two proxies are allocated with evidences $\pr{\tref{\tint}}$ and $\pr{\tref{\tbool}}$, respectively, both pointing to the same heap cell.  
Consequently, the assignment at line~3 and the dereference at line~4 both succeed, producing the final result $\true$.  
The diagram on the right of Figure~\ref{fig:mono-vs-guarded} illustrates this contrast: the left-hand side shows the monotonic case, where type updates are propagated directly in the heap, while the right-hand side shows the guarded case, where proxies mediate all accesses to the underlying memory cell.

Importantly, monotonic references satisfy the gradual guarantees~\cite{siekAl:snapl2015}, thereby still enabling fine-grained migration between static and dynamic typechecking, despite being more restrictive than guarded references. Furthermore, \citet{siekAl:esop2015} explain how one can recover flexibility on top of monotonic references via a syntactic discipline that consists of always assigning values at the unknown type, and deferring more precise type ascriptions to the client of the reference. 

Grift generalizes the idea of monotonic references to vectors and tuples~\cite{almahallawi:phd}.
Grift supports both guarded and monotonic semantics via a compiler flag. Similarly, GrEv features both semantics for data structures (\cref{tab:values}). A guarded reference is a proxy that consists of both the evidence attached to the reference and the pointer to the actual data (val\_ptr). Upon a more precise ascription, a new proxy is created with the new evidence, and a copy of the data pointer. In contrast, a monotonic reference holds the evidence and the actual data, and upon a more precise ascription, the evidence is updated in place. Therefore, monotonic references in GrEv enjoy benefit {\em (a)} above, namely avoiding allocating proxies for each alias. However, \citet{siekAl:esop2015} clarify that to enjoy benefit {\em (b)} above, it is necessary to recursively propagate updates to heap cells in depth, to adequately deal with nested structures. This process per se can be potentially costly and requires special care to avoid cycles. 
In contrast to the deep updates of monotonic references~\cite{siekAl:esop2015}, GrEv adopts a shallow approach, meaning that updated evidence information is not recursively propagated to nested structures. This means that GrEv cannot enjoy benefit {\em (b)} for monotonic references. (We compare all these different semantics in the performance evaluation in~\cref{sec:performance}).

Vectors are similar to references, except that they store their length---given that vector types do not carry length information---and hold (pointers to) multiple values.  
Tuples need not carry their length, as the information is available in the associated evidence.
For variants, GrEv assigns a unique identifier to each variant type, and one to each constructor; the type identifier is stored in the evidence (ev) associated to each value, and the constructor identifier (ctr\_id) is stored in the value itself (\cref{tab:values}). Given that variant types are in essence new atomic types, their evidence is fully precise and hence never changes, independently of the type at which they are accessed.


\paragraph{Closures} We observe that the discussion about guarded vs. monotonic data structures applies similarly to closures: one can either attach the evidence to the closure and let it evolve monotonically, or go through a proxy in order to allow different aliases to the same closure with different type constraints. This observation is, to the best of our knowledge, novel: \citet{siekAl:esop2015} do not explore the monotonic semantics for first-class functions, as they are modeled via substitution, not closures; and Grift only implements guarded closures. 

To briefly illustrate, with monotonic closures, an untyped identity function can be used with an argument of any type, as long as the closure itself has not been ascribed to a precise type incompatible with its use. Below on the left, \lstinline|f| is used with both integer and boolean arguments:
\begin{lstlisting}[numbers=none]
let f : ?->? = fun x -> x in        let f : ?->? = fun x -> x in
print_int(f 10)                     let g : int->int = f in
print_bool(f true) (* ok *)         print_int(g 10) ; print_bool(f true) (* fail *)
\end{lstlisting}
But on the right-hand side, the closure (aliased by both \lstinline|f| and \lstinline|g|) has monotonically evolved to a \lstinline|int->int| function because of the ascription of \lstinline|g|, so the application of  \lstinline|f| to \lstinline|true| fails at runtime. 

To assess the performance impact of monotonic closures, GrEv supports both closure semantics. \cref{tab:values} shows the memory representation of closures: the evidence, which includes the arity, the code label (fun\_ptr), which refers to the compiled code of the body of the function, produced during closure conversion, and the environment (either directly in the case of monotonic, or via a pointer in the case of guarded).





\paragraph{Semantic modes} In summary, Grift supports two semantic modes: guarded (Grift/G), with proxied data structures and proxied closures, and monotonic structures (Grift/MS), with unproxied data structures and proxied closures.   
For comparison, GrEv supports three modes: GrEv/G, with proxied data structures and closures; and two novel modes, GrEv/MV (\emph{monotonic values}), with unproxied data structures and closures, and GrEv/MC (\emph{monotonic closures}), which retains proxied data structures but uses unproxied closures.  
We include GrEv/MC specifically to isolate and measure the performance impact of monotonic closures (\cref{sec:performance}).

\subsection{Compiling with Evidence: Elaboration and Reduction}
\label{sec:elab-eval}

\begin{figure}[t]
 \includegraphics[width=\linewidth]{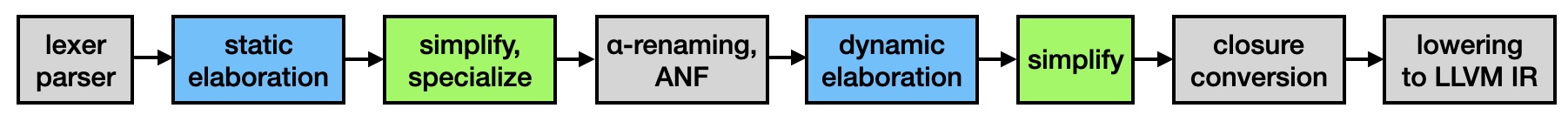}
\caption{The GrEv compilation pipeline. Standard phases are in grey, elaboration phases that introduce evidence ascriptions are in blue, and gradual typing optimizations are in green.}
\label{fig:pipeline} 
\end{figure}

The evidence-based formal semantics with AGT consists of an elaboration phase that inserts ascriptions, followed by the small-step semantics for reduction. When deriving a compilation strategy, the runtime semantics must instead be realized by a dynamic elaboration phase. The GrEv compilations pipeline is presented in \cref{fig:pipeline}: here we briefly discuss the two elaboration phases (blue boxes), and discuss the subsequent optimizations phases (green boxes) in \cref{sec:optims}.

\paragraph{Static elaboration}
As mentioned in \cref{sec:ev-semantics}, ascriptions are inserted everywhere consistency is used, aside from ascriptions already present in the source.  
This static elaboration phase (\cref{fig:pipeline}) affects values, function applications, data-structure accesses, and primitive operations, as illustrated below:

\begin{center}
\begin{minipage}[c]{0.30\linewidth}
\begin{lstlisting}[numbers=none]
let f = fun (x:?) -> x * 2 
in print_int (f false)
\end{lstlisting}
\end{minipage}
\hfill
\raisebox{0pt}[0pt][0pt]{$\overset{\text{elab}}{\leadsto}$}
\hfill
\begin{minipage}[c]{0.62\linewidth}
\begin{lstlisting}[numbers=none]
let f = <?->int>(fun (x:?) -> <int>x * <int>(<int>2)) in 
print_int <int>(<?->int>f <bool>(<bool>false))
\end{lstlisting}
\end{minipage}
\end{center}

Beyond the ascriptions of the number, boolean, and function values, each operand of the multiplication are also ascribed.  
Likewise, both the function and its argument are ascribed to $\tfun{\?}{\tint}$ and $\?$ respectively, yielding the evidences shown above.  
The primitive operation \lstinline|print_int| is treated in the same way: its argument must be ascribed to $\tint$ before the call.
Many of these ascriptions are redundant and can be eliminated (\cref{sec:optims}).


\paragraph{Dynamic elaboration}
The reduction rules of the evidence-based semantics (\cref{sec:ev-semantics}) introduce {\em dynamic} ascriptions, \ie~ascriptions to evidences that are determined at runtime. Consider the reduction rule for function application:
\begin{align*}
      \ev[1] (\lambda x. t)~\ev[2] u 
      \red& \evcolor{\cod(\ev[1])} (t[(\evcolor{\ev[2] \trans{} \dom(\ev[1])})u/x]) & \text{if defined, otherwise } \error
\end{align*}
The rule stipulates that the argument $\ev[2]u$ should be ascribed with the {\em domain evidence} of the function, $\dom(\ev[1])$, which results in computing consistent transitivity between both evidences, yielding either an error if inconsistent, or yielding a new evidence that is then used to decorate $u$ when substituted in the function body. Likewise, the resulting body is ascribed the {\em codomain evidence} of the function,  $\cod(\ev[1])$.

GrEv implements this semantics in the dynamic elaboration phase (\cref{fig:pipeline}) by ascribing the argument to the domain evidence at the call site, determining whether consistent transitivity holds prior to \texttt{call}ing the function. The ascription to the codomain evidence is inserted in the function body itself, checking consistent transitivity prior to \texttt{return}ing from the function call. The previous program---after some transformations (\cref{fig:pipeline})---is elaborated as follows:

\begin{lstlisting}[numbers=none]
let f = (fun (x:?) -> let x_asc = <int>x in 
                      let res = x_asc * <int>2 in
                      ascribe(res, cod(self)) in
let arg_0 = ascribe(<bool>true, dom(0, f)) in
print_int <int>(f arg_0)
\end{lstlisting}
The \lstinline|ascribe| form is handled during the lowering phase: it combines the evidence associated to its first argument with the evidence passed as its second argument, raising an error if consistent transitivity is undefined, or returning a new evidence value otherwise.
Here we use \lstinline|cod(self)| and \lstinline|dom(0, f)| to extract the codomain and (first) domain evidences from the closure evidence (\lstinline|self|, \lstinline|cod|, and \lstinline|dom| are also handled during lowering). It is important to highlight that this information can only be extracted at runtime. 
%
Other elimination forms, such as accesses to data structures, are treated similarly.

\subsection{Basic Gradual Typing Optimizations}
\label{sec:optims}

GrEv supports a couple of basic gradual typing optimizations, also implemented in Grift, which we briefly describe and illustrate.

\paragraph{Elaboration simplification} The formal definition of the elaboration uniformly inserts ascriptions with initial evidences, and can result in chains of ascriptions, as well as ascriptions that are unnecessary. GrEv reduces any chain of ascriptions to a single ascription with the combined evidence. It also eliminates an ascription when the evidence of the source is already at least as precise as that of the ascription itself, given that evidence composition is monotone. For example (including ascriptions for clarity): $\text{\lstinline{false::?}} \overset{\text{elab}}{\leadsto} 
\text{\lstinline{<bool>(<bool>false::bool)::?}} \overset{\text{simpl}}{\leadsto} \text{\lstinline{<bool>false::?}}$. The running example is then simplified as follows:

\begin{center}
\begin{minipage}[c]{0.28\linewidth}
\begin{lstlisting}[numbers=none]
let f = fun (x:?) -> x * 2 
in print_int (f false)
\end{lstlisting}
\end{minipage} 
\quad\raisebox{0pt}[0pt][0pt]{$\overset{\text{elab}}{\leadsto}$} ... \raisebox{0pt}[0pt][0pt]{$\overset{\text{simpl}}{\leadsto}$} \quad
\begin{minipage}[c]{0.55\linewidth}
\begin{lstlisting}[numbers=none]
let f = <?->int>(fun (x:?) -> <int>x * <int>2) in 
print_int (f <bool>false)
\end{lstlisting}
\end{minipage}
\end{center}
Besides the evidence on values, the only relevant evidence is the one next to \lstinline|x|, since its type is \lstinline|?|.


\paragraph{Specialized ascriptions}
Any elimination form (application, projection, etc.) of a term of the unknown type requires ascribing the term to the corresponding {\em germ}~\cite{lennonAl:toplas2022}, \ie~the least-precise type that admits the elimination form (\eg~\lstinline|?->?| for unary functions, \lstinline|vec[?]| for vectors). For instance:

\begin{align*}
\text{\lstinline{fun (x:?, y:?) -> x y}} 
\overset{\text{elab \& simpl}}{\leadsto} &
\text{\lstinline{fun (x:?, y:?) -> <?->?>x y}} \\
\overset{\text{specialize}}{\leadsto} &
\text{\lstinline{fun (x:?, y:?) -> (checkfun x) y}} 
\end{align*}

GrEv specializes the ascription to the germ \lstinline|?->?| to code that directly checks the top-level type constructor (written \lstinline|checkfun| above), which will be available at runtime in the evidence value bound to \lstinline|x|, without either updating the evidence or creating a proxy. This optimization is particularly effective on fully-untyped code, as it makes it possible to end up performing the same top-level tag checks that a safe dynamic language needs to do, but not more. In particular, in the guarded semantics, it avoids creating unnecessary (costly) proxies.


\paragraph{Unnecessary dynamic ascriptions}
Both source ascriptions and those introduced by elaboration are {\em static} in the sense that their target type is statically known. In contrast, reduction introduces {\em dynamic ascriptions}, whose target type is only known at runtime. For instance, as explained in \cref{sec:elab-eval}, when evaluating a function application, the argument is ascribed to the domain evidence of the closure being applied, and likewise, the return value is ascribed to its codomain evidence. The compiler does not know ahead of time the exact types that will be obtained from the closure evidence at runtime, but given that evidence evolves monotonically, the static information available denotes {\em upper bounds}. Exploiting these upper bounds makes it possible to determine that some dynamic ascriptions are unnecessary. In particular, this optimization is extremely effective on statically-typed code, because then the upper bound is a static type (\ie~fully precise), so dynamic ascriptions can be avoided altogether, yielding efficient code without runtime type checks.

For instance, consider an application \lstinline|f 1|, where \lstinline|f| has type 
\lstinline|int->int|. The elaboration yields \mbox{\lstinline|f ascribe(<int>1, dom(0,f))|}.
Here, the dynamic ascription is unnecessary, because \lstinline|<int>| cannot possibly gain further precision, so the application can be simplified to \lstinline|f <int>1|.


\paragraph{Optimized floats} 
Conversely to integers and booleans, float values need to use 64 bits, so it is not possible to tag them while preserving their immediateness. Dynamic languages therefore use boxed floats, allocated on the heap. In the case of gradual languages, uniformly using boxed floats is practical and simple, but misses the opportunity to use immediate floats in statically-typed code, thereby potentially hampering performance in float-intensive scenarios. 

After observing the best-case performance of Grift on float-intensive benchmarks, we noticed that, although not explicitly reported by~\citet{kuhlenschmidtAl:pldi2019}, Grift performs an optimization that consists in unboxing floats in statically-typed code.\footnote{Grift also performs a similar optimization on other primitive types, but the most noticeable effect is on floats.}
Unaware of any existing name for this optimization, here we call it the Dual Float Optimization (DFO). DFO consists in dealing with injections of an immediate float from $\tfloat$ into $\tunk$ as a boxing operation, effectively allocating the value on the heap, and dually, a projection from $\tunk$ to $\tfloat$ as an unboxing operation. 

 We have implemented DFO in GrEv, using evidence to track the typed-untyped boundaries for floats, performing boxing and unboxing as required. This ensures that a function that operates on statically-typed float values is compiled to code that directly uses float operations on immediate values. For instance, calling \lstinline|print_float x| where \lstinline|x:float| incurs no unboxing, while the same expression with \lstinline|x:?| generates an unboxing operation prior to applying the printing primitive. 
%

Unfortunately, DFO implies that some dynamic ascriptions that could be removed can no longer be optimized away, because they are needed to perform boxing and unboxing (and checking). Given this, and 
the lack of reported analysis of the effects of DFO in actual benchmarks, DFO in GrEv is optional, and we actually study its performance impact in \cref{sec:performance}.


\subsection{Comparison with Grift}
\label{sec:grev-grift}

GrEv and Grift compile essentially the same source language, although GrEv adopts an OCaml-like syntax, while Grift uses a Scheme-like syntax. Both compilers generally follow the same compilation strategy, for instance, by heap-allocating all data structures. The treatment of recursive types differs slightly as they are only used in GrEv for variant types, are isorecursive, while they are general and equirecursive in Grift~\cite{pierce:tapl}. 
Both compilers translate some closure applications into direct function calls using the techniques of \citet{keepAl:sfp2012}, but otherwise do not perform any general-purpose optimizations 
(\eg~type inference and specialization, constant folding, copy propagation, inlining, etc.).
Regarding optimizations of gradual typing per se, both GrEv and Grift perform similar optimizations, described in \cref{sec:optims}. 
GrEv is implemented in OCaml, and compiles to LLVM IR, while Grift is implemented in Racket and compiles to C, which is then compiled to LLVM IR. Therefore, both compilers enjoy the backend optimizations performed by LLVM. 
For memory management, GrEv and Grift use the same Boehm-Demers-Weiser conservative garbage collector, which implements a generational mark-sweep algorithm~\cite{boehmWeiser:spe1998,demersAl:popl1990}.

A first difference between Grift and GrEv is their handling of runtime type information, and the way checks are done, \ie~coercions vs evidence. As we have seen, in the simply-typed setting considered in this work, evidences are types. Coercions are different from types used in assertions, in that they represent actions (such as injection and projection, identity and failure, reading and writing from a structure, etc.).
A major difference between both compilers is the memory representation of values, which in GrEv is simpler and more uniform than in Grift. In Grift, values that have not been casted do not carry any type information, but once they are cast to the unknown type, they are represented as a pointer to a pair in the heap that consists of the underlying value and its type. This is the general principle of DFO explained above, but applied to all values, taking into account the specificities of each kind of values. In the case of booleans and integers the type information is stored in the 3 least significant bits and no reference is created. A closure in Grift can either be an actual closure or a proxy closure. An actual closure consists of a function pointer, a pointer to a function for casting the closure, and the environment captured by the closure. A proxy closure consists of a function pointer to a wrapper, a pointer to the underlying closure and a pointer to the coercion. 
When a closure is cast, Grift needs to discriminate at runtime which kind of closure it is. If it is an actual closure, Grift simply calls the caster function of the closure, which will create a new proxy closure. If it is a proxy closure, Grift composes the coercion from the cast with the coercion from the proxy closure and creates a new proxy, sharing the existing wrapper and underlying closure, but using the resulting coercion.

Performance-wise, the non-uniform value representation of Grift is beneficial on fully-typed configurations, where it can take advantage of the same representation it uses in the fully statically-typed variant of Grift, called StaticGrift~\cite{kuhlenschmidtAl:pldi2019}.
However, after a value is cast to a less precise type for the first time, a proxied version of that value is created and the advantage disappears. The experimental evaluation of the performance of both compilers (\cref{sec:performance}) sheds light on the impact of these representation choices.

\section{Performance Evaluation}
\label{sec:performance}

We now turn to the performance evaluation of GrEv. 
Before turning to the research questions directly related to the implementation of gradual typing with evidence, we first wish to ensure that the Grift and GrEv compilers are comparable. 
\begin{description}
\item[RQ1. Are the static baselines of Grift and GrEv comparable? (\cref{sec:rq1})] This is important to discard the possibility that the results that we observe from benchmarking both compilers on gradual programs are best attributed to  implementation differences unrelated to the management of gradual typing. 
\end{description}


Next, we settle to evaluate the overhead of gradual typing in GrEv, compared to Grift. We refine this general question as follows:

\begin{description}
\item[RQ2.] {\bf Is GrEv competitive with Grift in general? (\cref{sec:rq2})} 
We study this question by analyzing the performance of both compilers on (a)~fully-typed programs, (b)~fully-untyped programs, and (c)~partially-typed programs.
\end{description}

\begin{description}
\item[RQ3.] {\bf How do the different semantic modes affect performance? (\cref{sec:rq3})} 
In order to assess the impact of monotonic semantics, we run the benchmarks on the available implementations of Grift---Grift/G (with guarded semantics) and Grift/MS (with monotonic structures and guarded closures)---and those of GrEv: GrEv/G (guarded), GrEv/MC (monotonic closures, guarded structures) and GrEv/MV (all values monotonic). 
\end{description}

\begin{description}
\item[RQ4.] {\bf How effective is the dual-float optimization? (\cref{sec:rq4})} 
We consider GrEv with and without the dual-float optimization (DFO~\cref{sec:optims}), and compare it to Grift, using the guarded semantics of both compilers,  on float-intensive programs. 
\end{description}

\subsection{Experimental Methodology}

\paragraph{The Grift benchmark suite.} 
We reuse the benchmark suite from the Grift repository, which includes 8 benchmarks evaluated in the original Grift paper~\cite{kuhlenschmidtAl:pldi2019}, which only measures Grift with proxied data structures, in addition to 2 benchmarks that were added to the Grift repository to assess the performance impact of monotonic references~\cite{almahallawi:phd}. The 8 benchmarks come from a variety of sources, including the Scheme benchmark suite (R6RS), the PARSEC benchmarks, the Computer Language Benchmarks Game (CLBG), the Gradual Typing Performance Benchmarks (GTP), as well as two textbook algorithms (matrix multiplication and quicksort). The 2 additional benchmarks involve data structures. \cref{fig:suite} summarizes the benchmarks; the last four exercise floats.

\newcommand{\markbg}[1]{\setlength{\fboxsep}{0pt}\fcolorbox{lightgray}{lightgray}{#1}}
\begin{figure}
\begin{footnotesize}
\begin{tabular}{@{}llll@{}}
\toprule 
\textbf{Benchmark} &  \textbf{Source} & \textbf{Description} \\
\midrule
\bench{array} & Grift & iterative array generation and reversal in a recursive function\hfill \markbg{RQ1,2,3}\\
\bench{matmult} & textbook & triply-nested loop multiplicating matrices of integers \\
\bench{quicksort} & textbook & sorting integer arrays (worst case, \ie~already ordered input) \\
\bench{qsort-mpairs} & Grift & variant of quicksort, sorting arrays of pairs of integers  \\
\bench{sieve} &  GTP & compute prime numbers via Sieve of Eratosthenes, with streams (recursive type) \\
\bench{tak} & R6RS & triply recursive integer function (Takeuchi) \\
\midrule
\bench{blackscholes} & PARSEC & numerical computation of Black-Scholes partial differential equation 
\hfill \markbg{RQ1,4} \\
\bench{fft}  & R6RS & computes Fast Fourier Transform for real-valued points\\
\bench{n-body} & CLBG & computes the orbits of Jovian planets using symplectic integrator \\
\bench{ray}  & R6RS & Ray tracing a scene \\
\bottomrule
\end{tabular}
\end{footnotesize}
\caption{Benchmark suite.}
\label{fig:suite}
\end{figure}

\paragraph{Experimental setup.}
The experiments were conducted on a dedicated machine with an AMD Ryzen 9 9900X3D 24-core  processor at 4.4 GHz (128 MB L3 cache and 12MB L2 cache), and 123 GB of RAM running Debian Linux 13.2. 
The LLVM version is 19.1.7, Racket is version 8.18 (Chez Scheme). 
For Grift, as the available version in the Racket package manager fails to install on newer Linux distributions, we used a container to install Grift and compile the benchmarks, but ran the configurations outside the container.
We use a Docker image based on Ubuntu 20.04 to install the required dependencies for Grift and install a specific version of Grift from the available source code in GitHub (links for \href{https://github.com/Gradual-Typing/Grift.git#pldi19}{Grift/G} and for \href{https://github.com/Gradual-Typing/Grift.git#fix109}{Grift/MS}).
The Racket version used in this Docker image is 8.18 (Chez Scheme) and the version of Clang is 10.0.0. We attempted but failed to run Grift with newer versions of Clang.
All time measurements were made using real time, obtained from the internal timing utilities of each language.
The results we present correspond to the average of 10 runs for partially-typed configurations and the average of 50 runs otherwise.

\paragraph{Measuring the performance lattice.}
\citet{takikawaAl:popl2016} first observed that evaluating the performance of gradually-typed languages requires one to consider, for each program under evaluation, all the possible versions that can be obtained with more or less typing annotations. In the case of fine-grained gradual typing, as supported in \lang, all variable bindings are subject to having a type annotation or not, and even worse, the type annotation can be more or less precise. To tackle this exponential space explosion, we follow the linear sampling based approach of \citet{greenmanMigeed:pepm2018}, also used to evaluate Grift~\cite{kuhlenschmidtAl:pldi2019,almahallawi:phd}. 

Concretely, we reimplement the Grift ``dynamizer'' tool in order to work on the \lang syntax processed by GrEv. This tool takes a statically-typed program, a number of samples and a number of bins to be uniformly sampled. It localizes all type annotations in the program and generates gradual versions of those types, while ensuring that the generated configuration lies within the desired bin. To ensure randomness we perform a random weighted shuffle of the list of type annotations. The process repeats until all bins have the expected number of configurations.

For each benchmark we generate a number of configurations proportional to the number of static types in the original program.
Specifically, for a program with $n$ static type nodes we generate $10n$ configurations.
To ensure that the generated sample contains configurations across the different proportions of static types in the lattice, these configurations are divided among 10 bins.

\subsection{Static Baselines (RQ1)}
\label{sec:rq1}

\begin{figure}[t] 
\includegraphics[width=\textwidth]{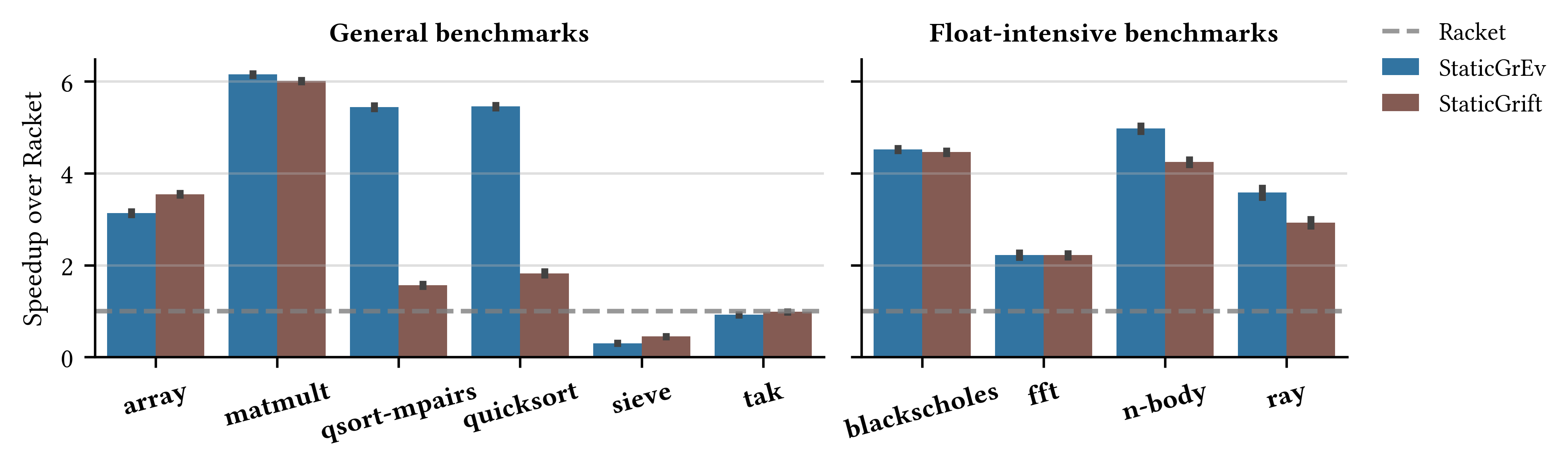}
\caption{Speedup of StaticGrEv and StaticGrift over Racket (higher is better).
General benchmarks are shown on the left and float intensive-benchmarks on the right.
}
\label{fig:static} 
\end{figure}

\cref{fig:static} reports the relative performance of StaticGrEv and StaticGrift compared to Racket.
Benchmarks are separated into general benchmarks (left) and float-intensive benchmarks (right).
The dashed horizontal line shows the performance of Racket, values above that line reveal better performance than Racket.

For general benchmarks, we observe that StaticGrEv and StaticGrift are overall similar.
The notable exceptions are \bench{qsort-mpairs} and \bench{quicksort} where StaticGrEv performs significantly better (speedups of over 5x on the former, and of almost 2x on the latter), and \bench{array} where StaticGrift performs slightly better.
Both compilers are better than or comparable with Racket on all benchmarks, except on \bench{sieve} where both compilers are slower than Racket.

On float-intensive benchmarks, once again we observe that both compilers are overall similar, with slightly better performance for StaticGrEv on \bench{n-body} and \bench{ray}.
Both compilers achieve better performance than Racket, with speedups ranging from 2x to over 4x.
Racket is at a disadvantage on float-intensive benchmarks because floats are heap-allocated, while the two static compilers take advantage of unboxed, immediate floats.

Despite different implementations both compilers achieve similar performance in 8 of the 10 benchmarks, with the notable exceptions being \bench{qsort-mpairs} and \bench{quicksort} where StaticGrEv is significantly faster than StaticGrift.

\newenvironment{rqbox}{%
\begin{mdframed}[roundcorner=10pt,skipabove=\baselineskip, skipbelow=\baselineskip]
\begin{tabular}{lp{32em}}
}{%
\end{tabular}
\end{mdframed} 
}

\begin{rqbox}
{\bf RQ1} & 
Despite their different implementations, the static baselines of both Grift and GrEv, namely StaticGrift and StaticGrEv, are comparable in most benchmarks.
\end{rqbox}

This result confirms that it makes sense to compare GrEv and Grift on gradual programs, as the performance differences between both compilers unrelated to gradual typing mechanisms are overall small.

\subsection{General Performance (RQ2)}
\label{sec:rq2}

We now use the general benchmarks to compare the performance of Grift and GrEv
in order to assess whether GrEv is competitive with Grift overall. 
In the benchmarks, we use GrEv and Grift across their different variants: GrEv/G, GrEv/MC, GrEv/MV, Grift/G and Grift/MS. This section concentrates on general observations regarding the two compilers; we discuss specific findings regarding the performance impact of the various monotonic semantics in \cref{sec:rq3}.

\begin{figure}[t]
\includegraphics[width=\textwidth]{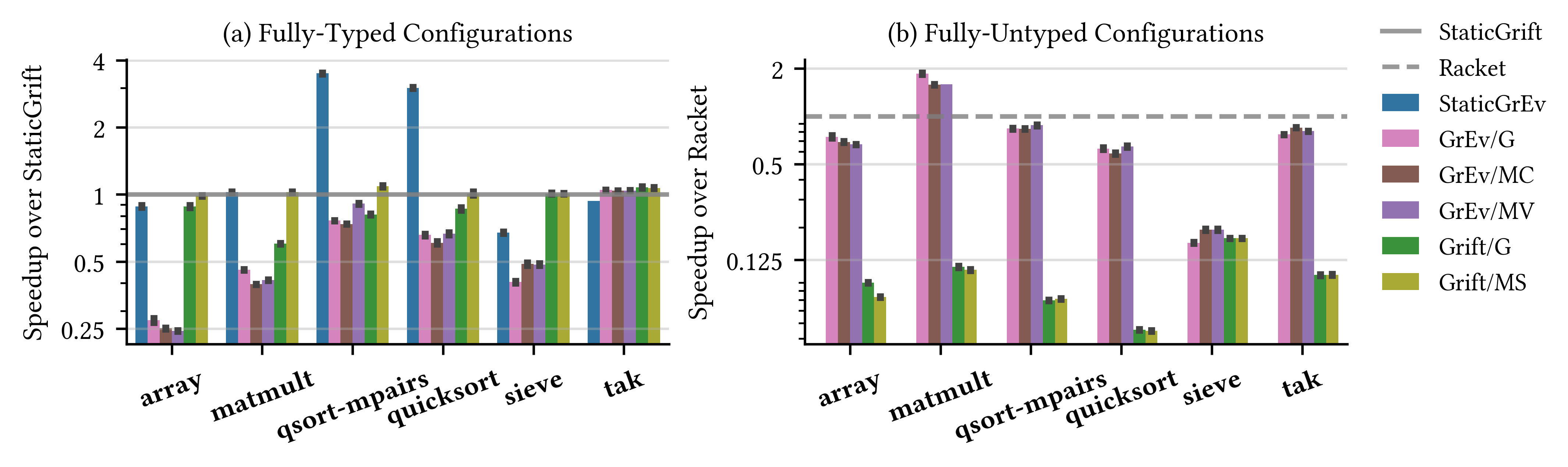}
\caption{Speedup of GrEv and Grift on fully-typed and fully-untyped programs (higher is better).\newline
The horizontal line shows the respective baseline performance:
(a) fully-typed configurations, speedup is over StaticGrift; 
(b) fully-untyped configurations, speedup is over Racket.
}
\label{fig:typed-untyped}
\end{figure}

\paragraph{Fully-typed programs} \cref{fig:typed-untyped}(a) shows the relative performance of both compilers on fully-typed programs, as the speedup over StaticGrift.
We observe that Grift achieves better performance than GrEv on most benchmarks.
On \bench{tak} both compilers have similar performance, approaching StaticGrift. 
In general, GrEv performs worse than StaticGrEv---with speedups ranging from 0.25x in \bench{array} to near 1x on \bench{tak}.
GrEv is not able to take full advantage of static types as well as Grift, and only achieves performance on par with StaticGrift in one of the benchmarks.
This seems to be related with the fact that, as described in \cref{sec:grev-grift}, Grift represents values that have not been cast similarly to how they are represented in StaticGrift, and only creates a proxy when a value is cast.
In contrast, GrEv uses a uniform value representation (\cref{tab:values}), so there is some overhead related to gradual typing even on fully-typed configurations.

\paragraph{Fully-untyped programs} \cref{fig:typed-untyped}(b) shows the relative performance of both compilers on fully-untyped programs, as the speedup over Racket.
As can be observed, GrEv performs significantly better than Grift on untyped programs. GrEv is competitive with Racket, while Grift is always at least 5x slower than Racket. 
Speedups for Grift range from 0.05x to 0.2x, while speedups for GrEv range from near 0.2x to over 1x---GrEv even outperforms Racket on \bench{matmult}.
The only exception is \bench{sieve}, where both compilers both perform similarly badly. 
Given that \bench{sieve} involves heavy use of recursion and closures, and that neither GrEv nor Grift optimize closures or perform general-purpose optimizations other than the ones done by Clang/LLVM, we conjecture that this bad performance might be due to the optimized Racket implementation of these features.

\begin{figure}[t]
\includegraphics[width=\textwidth]{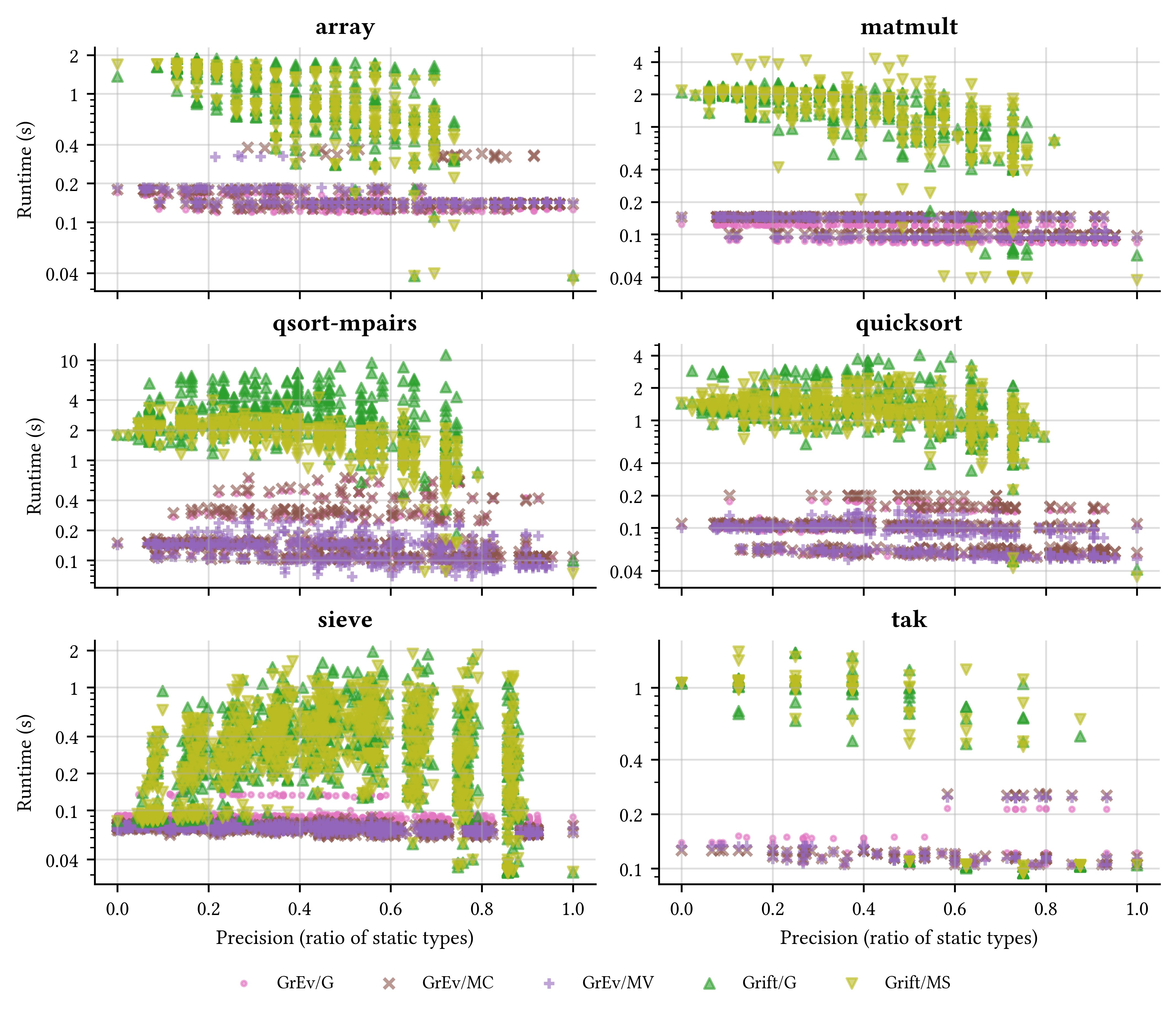}
\caption{
Performance of GrEv and Grift on partially-typed configurations.
The x-axis indicates the ratio of static types (0 = fully untyped, 1 = fully typed).
The y-axis shows runtime in seconds on a logarithmic scale (lower is better).
Each marker represents an individual configuration which was run 10 times.}
\label{fig:partial-dots}
\end{figure}

\begin{figure}[t]
\includegraphics[width=\textwidth]{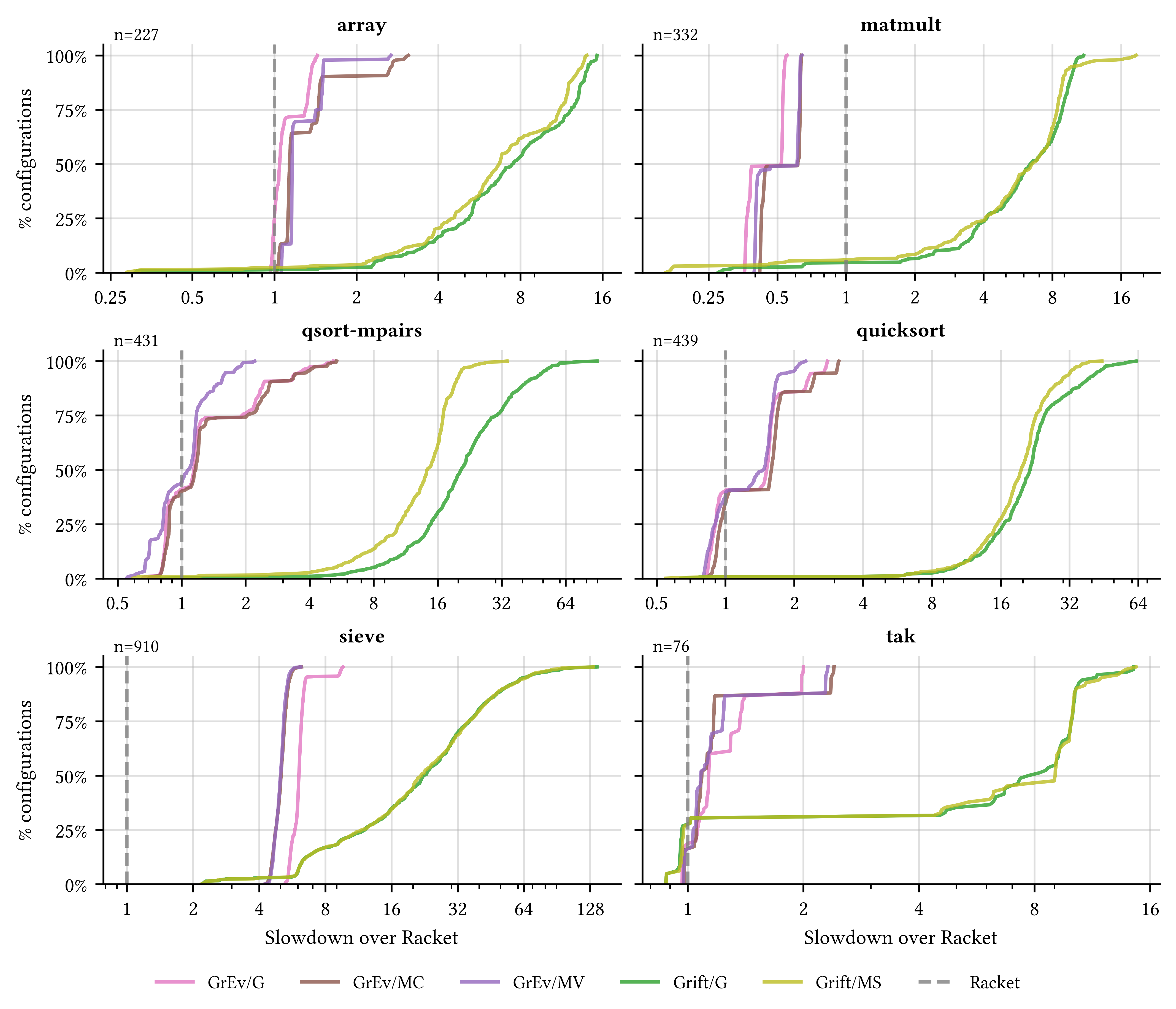}
\caption{Cumulative performance of GrEv and Grift on partially-typed configurations.
The x-axis represents the slowdown with respect to Racket on a logarithmic scale (left is better).
The y-axis is the percentage of configurations above that slowdown.
Each configuration was run 10 times.}
\label{fig:partial-lines}
\end{figure}

\paragraph{Partially-typed programs} 
In \cref{fig:partial-dots} we compare the performance of GrEv and Grift across their different variants, focusing on partially-typed configurations of the benchmarks.
Configurations are divided among 10 bins.
Each bin corresponds to an interval of the proportion of static types, ranging from \(0\%\) (0.0 on the left of the \(x\)-axis) to \(100\%\) (1.0 on the right).  
In addition, for each benchmark we manually include the fully-typed and fully-untyped configurations, which appear as the two extremes at 0.0 and 1.0 on the \(x\)-axis.


For partially-typed configurations we observe that GrEv performs better in general, except (as already observed above) on fully-typed configurations.
Most notably, the variation in performance across different configurations is smaller for GrEv than for Grift, indicating more predictable performance with GrEv.
In the case of GrEv we see ``flat profiles'', with groups of points concentrated near certain runtimes, forming horizontal lines across different ratios of static types.
This suggests that there are specific type annotations whose presence or absence makes a configuration experience a discrete increment or decrement in runtime, regardless of the ratio of static types for the whole program.
For Grift we observe some configurations with a ratio of static types above 0.5 having drastically better performance than the rest of configurations.%
\footnote{For Grift, some configurations appear to be missing just to the left of the fully-typed configuration.
This is related to how Grift generates configurations: we found that the Grift dynamizer reports precisions greater than one, which is naturally impossible. Upon manually inspecting the source code, we noticed that the Grift dynamizer was reporting incorrect values of precision, consistent with miscounting the number of static types by a constant specific to each benchmark. To correct this, we applied an adjustment factor that shifts the shown precision to the left.
On some benchmarks this results in an interval of precision where no configurations are generated.}

In \cref{fig:partial-lines} we compare the {\em cumulative} performance of GrEv and Grift across their different semantics on partially typed configurations. 
The $x$-axis is the relative slowdown over Racket in {\em logarithmic} scale and the $y$-axis is the percentage of configurations above that slowdown.
A vertical line would indicate identical performance for all configurations.
The closer to a vertical line the more predictable the performance is across configurations.
Near horizontal lines indicate discrete increases in runtime from one set of configurations to another.

This visualization confirms that GrEv generally performs significantly better than Grift over the whole spectrum, except that Grift performs better on (close to) fully-typed configurations. For the vast majority of configurations, GrEv outperforms Grift by a large factor (recall that the $x$-axis is in logarithmic scale). A notable example is \bench{matmult} where
GrEv performs on average twice as fast as Racket, while Grift is slower (up to 16x) than Racket on more than 95\% of the configurations. 
The performance of GrEv is also more predictable: there are much smaller differences between its best-case performances and worst-case performances.
For instance, on \bench{sieve}, the best and worst case slowdowns of Grift are around 2x and 128x, respectively, while
for GrEv/G, the performance window ranges from 5x to 9x.

Both compilers exhibit ``discontinuities" (near horizontal segments) in some benchmarks: \bench{qsort-mpairs} and \bench{quicksort} for GrEv, and \bench{tak} for both GrEv and Grift.
This corresponds to the vertical gaps we see in \cref{fig:partial-dots}, where there are no configurations within that vertical area.
This suggests that there are static types in the source program whose presence has an important performance impact on the overall program performance.
Ideally, one wants to avoid these cases as this means that the programmer changing a single type annotation may result in drastically different performance, confirming observations in prior work, and the importance of research on migrational typing~\cite{camporaAl:popl2018,camporaAl:icfp2018} to assist programmers in navigating the spectrum.
In this regard, the fact that GrEv is noticeably more predictable than Grift is a strong argument in favor of the evidence-based approach of GrEv. Also, it is important to observe that these discontinuities tend to manifest for a much smaller fraction of configurations in GrEv than in Grift (visually, the discontinuities occur closer to the top, as can be seen particularly clearly in \bench{sieve} and \bench{tak}).

\begin{rqbox}
{\bf RQ2} & GrEv performs significantly better than Grift overall, except on fully-typed programs, and is more predictable than Grift across the type precision spectrum.
\end{rqbox}
This result highlights that aiming to reduce to a maximum the overhead on fully-typed programs (using non-uniform value representations) is in tension with general performance and stability across the type precision spectrum. 
The performance analysis of DFO below (\cref{sec:rq4}) dives into this issue in the specific case of floats.


\subsection{Semantic Modes (RQ3)}
\label{sec:rq3}

\paragraph{Fully-typed programs} 
The results presented in \cref{fig:typed-untyped}(a) show the benefits of monotonic references in Grift, as Grift/MS performs better than Grift/G, achieving close to 1x speedup on all benchmarks, while Grift/G exhibits some noticeable slowdowns in \bench{array}, \bench{matmult}, \bench{qsort-mpairs} and \bench{quicksort}.
Between the three semantics of GrEv, we observe small differences in performance, with GrEv/G being slightly faster on \bench{array} and \bench{matmult}, GrEv/MV being slightly faster on \bench{qsort-mpairs} and both monotonic versions being slightly faster on \bench{sieve}.
Overall Grift obtains a greater benefit from monotonic structures on fully-typed programs.
This is expected as Grift implements deep monotonic structures.
This means that when a vector is cast to a static type, Grift casts all the values in that vector, and from that point forward it does not need to perform casts when reading or writing from that vector.
In contrast, shallow monotonic structures in GrEv cannot avoid checks when reading or writing from that vector---they only allow avoiding new proxy allocations upon type ascriptions.

\paragraph{Fully-untyped programs} Looking back at \cref{fig:typed-untyped}(b), one can observe that the monotonic semantics have little to no impact on fully-untyped programs. 
This is not particularly surprising as monotonic semantics are meant to exploit static type information, which is absent in this scenario.
Specifically, GrEv/G performs slightly better than the other two semantics in \bench{matmult} and slightly worse in \bench{sieve}.
Similarly, performance is similar for both semantics of Grift, with Grift/G performing slightly better in \bench{array}.

\paragraph{Partially-typed programs}
On partially-typed configurations (\cref{fig:partial-lines}) the benefits of the monotonic semantics do manifest, to some extent.
In Grift we observe significant improvements in \bench{array}, \bench{qsort-mpairs} and \bench{quicksort}, while the performance is similar in the rest of the benchmarks, except in \bench{matmult} where there is a significant slowdown in the worst performing configurations.
For GrEv we see significant improvements in \bench{qsort-mpairs}, \bench{quicksort} and \bench{sieve}.
Notably, in \bench{qsort-mpairs} both GrEv/G and GrEv/MC experience a slowdown of over 5x while the worst slowdown for GrEv/MV is less than 2x.
However, on some benchmarks the monotonic semantics perform worse than the guarded semantics.
For GrEv, the monotonic versions underperform in \bench{array}, \bench{matmult} and \bench{tak}.

One benchmark where GrEv/MV underperforms is \bench{array}.
This benchmark involves the use of loops and mutually recursive functions to generate and reverse integer arrays.
In order to understand this surprising result, we profiled the worst-case configuration and compared the results for the three semantics of GrEv, with and without Clang/LLVM optimizations. We observed that the difference was insignificant when such optimizations were disabled, suggesting that some aspect of GrEv/G is more amenable to optimizations in this benchmark.

Depending on the benchmark we observe GrEv/MC to be close either to GrEv/G (\bench{qsort-mpairs}, \bench{quicksort}) or to GrEv/MV (\bench{matmult}, \bench{sieve}, \bench{tak}).
This suggests that the difference between GrEv/G and GrEv/MV is sometimes due to monotonic closures and sometimes due to monotonic structures.
Both \bench{sieve} and \bench{tak} make heavy use of recursion and do not use arrays so it is unsurprising for GrEv/MC to be close to GrEv/MV.
Likewise, \bench{qsort-mpairs} and \bench{quicksort} make heavy use of arrays, so it is expected for GrEv/MC to be closer to GrEv/G on those benchmarks.

\begin{rqbox}
{\bf RQ3} & 
On partially-typed configurations both compilers benefit from the monotonic semantics, even though they can perform similarly or slightly worse in some benchmarks. This also applies to monotonic closures, which effectively complement monotonic structures in certain scenarios.
\end{rqbox}

\subsection{Dual-Float Optimization (RQ4)}
\label{sec:rq4}

\begin{figure}[t]
\includegraphics[width=\textwidth]{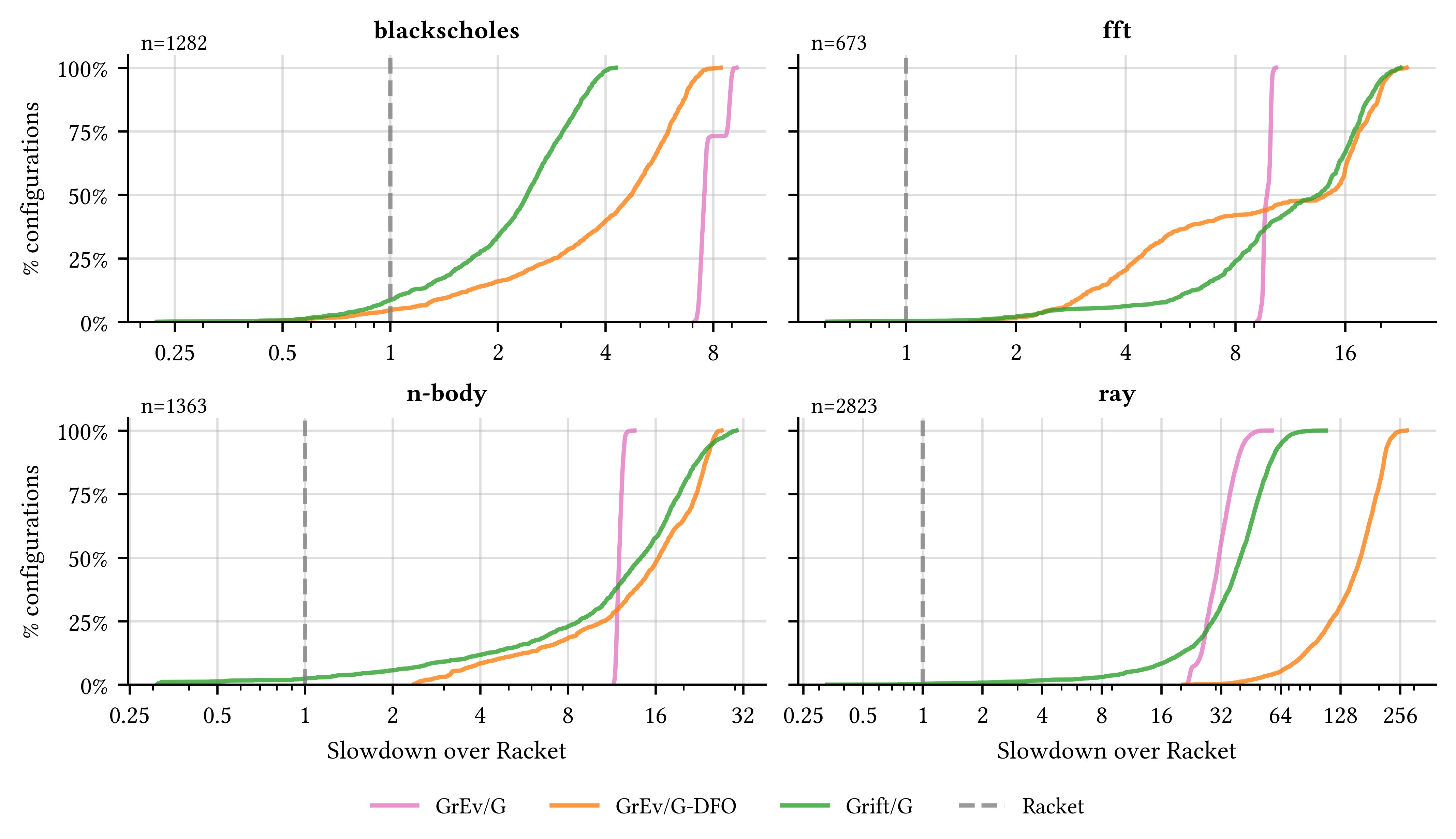}
\caption{Cumulative performance of GrEv with and without Dual Float Optimization (DFO), and Grift on partially-typed configurations of the floating point benchmarks.
The x-axis represents the slowdown with respect to Racket (on a logarithmic scale).
The y-axis is the number of configurations above that slowdown.}
\label{fig:dfo}
\end{figure}

In order to better understand the impact of the dual float optimization (DFO) described in \cref{sec:optims}, we report on the comparison between Grift (with DFO) and GrEv with and without DFO.
In \cref{fig:dfo} we compare the cumulative performance of GrEv/G (no DFO), GrEv/G-DFO, and Grift/G on the floating point benchmarks (\cref{fig:suite}).

We observe that DFO can enable substantial performance improvements as is observed in \bench{blackscholes} for all configurations and in \bench{fft} and \bench{n-body} for the best performing configurations.
After adding DFO the slowdown of the 25\% best performing configurations goes down from over 7x to under 3x for \bench{blackscholes} and from over 9x to close to 4x for \bench{fft}.
In \bench{fft} and \bench{n-body} these improvements occur in almost-fully-typed configurations.

With both compilers, some configurations perform much worse than Racket.
In particular, in \bench{ray} the three implementations have quite dramatic worst-case slowdowns, with GrEv/G-DFO having a worst-case slowdown over 256x, Grift over 70x, and GrEv/G around 60x.

GrEv/G demonstrates significantly more predictable performance than the implementations with DFO, with nearly vertical lines across all four benchmarks, and better median performance on three of the four benchmarks (\bench{blackscholes} being the exception). In contrast, both Grift/G and GrEv/G-DFO show substantial variability between best-case and worst-case performance. The most dramatic example is \bench{ray}, where Grift/G ranges from less than 0.4x slowdown to more than 64x slowdown, while GrEv/G-DFO ranges from less than 32x to over 128x slowdown.

Regarding best-case performance the best implementation is Grift/G. It achieves good best-case performance on all benchmarks, surpassing Racket in three of the four benchmarks (\bench{blackscholes}, \bench{n-body} and \bench{ray}), but its worst-case performance is worse than GrEv/G on all benchmarks except \bench{blackscholes}.

GrEv/G-DFO does not perform better than GrEv/G in all the floating point benchmarks.
While in \bench{blackscholes} it achieves better performance in all configurations, in other benchmarks the improvement is not so clear.
On \bench{fft} and \bench{n-body} it achieves better best-case performance, but the worst-case performance is significantly worse.
Furthermore, in \bench{ray} it performs worse than GrEv/G on the vast majority of configurations.

We identify two factors that likely contribute to these performance regressions.
First, as described in \cref{sec:optims}, GrEv/G-DFO is unable to perform some optimizations that GrEv/G can, such as removing unnecessary dynamic ascriptions.
Second, implementing DFO necessitates handling the different runtime representations of floats.
This makes the implementation of consistent transitivity more complicated, which incurs additional runtime overhead.
On some benchmarks these factors can outweigh the benefits of DFO, resulting in performance degradation.
Also, with DFO, when a value of type $\tfloat$ is ascribed to $\tunk$, a new heap-allocated float needs to be created---without DFO an ascription to  $\tunk$ is free.
This means that depending on where the typed-untyped boundaries occur, DFO can perform many memory allocations that are not needed without DFO.
On the other hand, on programs with many operations on values known to be floats, DFO avoids having to unbox float values, and avoids allocating new floats on the heap.

\begin{rqbox}
{\bf RQ4} &
The benefit of DFO in GrEv is mixed, it provides better best-case performance on some benchmarks, but worse worst-case performance.
Grift beats GrEv in best-case performance, but has worse worst-case performance in three out of the four benchmarks.
GrEv/G is more predictable than GrEv/G-DFO and Grift/G.
\end{rqbox}

This analysis of the performance impact of DFO tends to confirm the observation that using different value representations in order to boost the efficiency of statically-typed code can yield rather high performance degradation in partially-typed configurations. Exploiting program analyses, including dynamic ones, in order to better choose which representation to use might be a worthwhile venue to explore.



\subsection{Threats to Validity}
\label{sec:threats}

Both Grift and GrEv are fairly straightforward compilers for a small language, \lang. 
The results obtained here may not generalize when dealing with a more complex language, for instance feature parametric polymorphism, record subtyping, and more. The 
Grift benchmarks were handpicked by \citet{kuhlenschmidtAl:pldi2019} from other benchmark suites, selected based on the features covered by \lang. A consequence of this is that several benchmarks deal with arithmetic, and overall, the results could vary on other types of benchmarks.

Another threat to validity is that the differences observed between GrEv and Grift could come from implementation differences unrelated to gradual typing, for instance the fact that Grift compiles to C, while GrEv goes straight to LLVM IR, or the fact that we had to use a slightly older version of Clang to run Grift than to run GrEv. As mentioned in~\cref{sec:grev-grift}, Grift and GrEv also have slightly different ways of handling recursive types, but only one benchmark (\bench{sieve}) makes use of these types. Overall, our first research question~(\cref{sec:rq1}) addresses this threat, showing that StaticGrift and StaticGrEv exhibit overall similar performance on these benchmarks.

While both compilers perform essentially the same optimizations, \citet{almahallawi:phd} reports using hashconsing in Grift/MS in order to reuse and efficiently compare structural types, although they do not report on the specific impact of the technique. At present, we have not implemented hashconsing for evidence in GrEv, but we hope to assess its effectiveness in the near future. 

Regarding the evaluation approach, the fact that both compilers use syntactically different source languages implies some threats to validity as well. We had to manually translate the Grift benchmarks to the GrEv source language, and it is possible for small differences in the translation of the benchmarks to result in performance differences. This threat was mitigated by paying great care to keep the translations as close as possible to their Grift counterparts, and by checking that they indeed compute the same results. Also, while the partially-typed configurations are generated by a similar sample generation algorithm, they are not identical; it is possible that the generated samples favor one implementation over the other. However, we mitigate this risk by generating numerous samples for each bin of precision (ratio of static types), making it very unlikely that the general results are affected by these individual differences.

Finally, this work only evaluates the impact of gradual simple types, and the results might not transpose to more complex types. Evaluating evidence-based compilation of more advanced gradual typing disciplines is not addressed in this work. Grift currently only supports gradual simple types as well. The AGT literature has explored different notions of evidence and consistent transitivity for a wide range of typing disciplines, however, only proof-of-concepts interpreters have been developed so far.
These developments could now be studied in the context of GrEv, opening the door to new implementation challenges.

\section{Related Work}
\label{sec:related}

Research on gradual typing spans several axes, from semantic foundations to compiler engineering and performance optimization.  
We organize this discussion around four major themes: (i) Abstracting Gradual Typing (AGT) and its applications, (ii) implementations and performance studies, (iii) runtime representations and optimization techniques, and (iv) nominal approaches and other approaches for efficiency.

\emph{AGT foundations and evidence-based semantics.}
The Abstracting Gradual Typing (AGT) framework~\cite{garciaAl:popl2016} provides a systematic method to derive gradual type systems via abstract interpretation~\cite{cousot:popl1977}.  
This design has been extended to many typing disciplines, including effects~\cite{banadosAl:jfp2016}, refinement types~\cite{lehmannTanter:popl2017,vazouAl:oopsla2018}, set-theoretic types~\cite{castagnaLanvin:icfp2017,toroTanter:sas2017}, security typing~\cite{toroAl:toplas2018}, parametric polymorphism~\cite{toroAl:popl2019,labradaAl:jacm2022,labradaAl:oopsla2022}, flexible data types~\cite{malewskiAl:oopsla2021}, probabilistic programming~\cite{yeAl:oopsla2023}, and recently, sensitivity typing~\cite{arquezAl:csf2025}.  
The use of AGT has remained primarily theoretical; apart from prototype interpreters,
the efficient implementation of the evidence-based runtime model has not been explored until now. This work closes this gap by mapping evidence-based semantics to low-level representations and evaluating its efficiency relative to the Grift coercion-based implementation.

\emph{Implementations and performance studies.}
Early practical implementations of sound gradual typing exposed severe performance limitations.  
\citet{takikawaAl:popl2016} showed, through an extensive evaluation of Typed Racket, that the runtime checks introduced by sound gradual typing could slow programs by several orders of magnitude.  
This result motivated a sustained effort to make sound gradual typing efficient.
A key milestone is the Grift compiler~\cite{kuhlenschmidtAl:pldi2019}, the first ahead-of-time compiler for a gradually-typed language with structural types and fine-grained, sound semantics.  
Grift supports both type-based casts and space-efficient coercions~\cite{hermanAl:hosc10}, normalizing chains of coercions at runtime to control both space and time overhead.  
It supports two runtime modes: \emph{guarded} semantics, which attach coercions to proxies to heap values, and \emph{monotonic references}~\cite{siekAl:esop2015}, which enforce runtime type information to monotonically evolve toward more precise types. \citet{almahallawi:phd} confirms that monotonic references offer significant speedups without compromising soundness. Our compiler, GrEv, builds upon some of these ideas but replaces coercions with evidence, and explores the use of monotonicity to closures, not only data structures. 

\emph{Runtime representations and optimization.}
Space efficiency and the reduction of redundant checks have been central goals in improving the performance of gradual typing.  
\citet{hermanAl:hosc10} introduced \emph{coercions} as compact and composable representations of runtime casts.  
\citet{toroTanter:scp2020} later demonstrated that, for mutable references, coercions and evidence are equivalent in both expressivity and compactness.  
They also showed that an evidence-based semantics can be made space-efficient by adopting similar nonstandard reduction rules. Subsequently, \citet{tsudaA:ecoop2020} proposed a \emph{coercion-passing style} to eliminate the need for such nonstandard reductions by making coercions first-class citizens. They implement this approach in Grift, successfully preventing stack overflows, although at the cost of execution times up to three times higher.  
In GrEv, evidence values are already compact; however, the current implementation does not prevent stack overflows.  
Exploring and evaluating a similar evidence-passing style is an interesting venue for future work.

Beyond coercion-based implementations, \citet{vitousekAl:dls2014,vitousekAl:popl2017} introduced \emph{Reticulated Python}, a gradually-typed variant of Python based on the \emph{transient} semantics.  
Transient semantics performs only shallow, first-order checks at type boundaries and ensures \emph{open-world soundness}, meaning that gradual code can safely interact with untyped libraries. \citet{greenmanFelleisen:icfp2018} analyze the different soundness results obtained from the traditional guarded semantics, the transient semantics, and the erasure semantics, and evaluate their performance in Typed Racket~\cite{tobinFelleisen:popl2008}. The erasure semantics consists in relying on the underlying dynamic language checking mechanisms, and is therefore not applicable in our setting.

\emph{Nominal gradual typing and structural extension.}
\citet{muehlboeckTate:oopsla2017} show that adopting a nominal typing discipline, rather than a structural one, can yield substantially more efficient runtime representations for sound gradual typing.
They formalize a nominal, object-oriented core language. 
In their design, function types are represented through interfaces and closures through classes.  
Their evaluation uses benchmarks from \citet{takikawaAl:popl2016} and \citet{vitousekAl:popl2017}, and compares performance against both C\#~\cite{biermanAl:ecoop2010} and Reticulated Python.  
\citet{muehlboeckAndTate:oopsla21} later extended this line of work with support for dynamically-typed structural constructs such as records and lambda expressions.

\emph{Other optimizations.}
\citet{rastogi:popl2012} describe an inference-based approach to eliminate some runtime casts while preserving semantics. 
This approach was adapted by \citet{vitousekAl:dls2019} to optimize transient gradual typing in Reticulated Python. Their evaluation on a suite of Python benchmarks showed that partially-typed programs can achieve performance close to their untyped counterparts. 
Recently, \emph{discriminative typing}~\cite{camporaAl:popl2024} introduces type-based optimizations that produce separate fast and slow versions of functions depending on the available type information at the call site, similarly to the approach explored by \citet{allendeAl:dls2013} in the context of Gradualtalk~\cite{allendeAl:scp2014}.  
These optimizations are orthogonal to the underlying runtime mechanism—whether based on coercions, casts, or evidence—and could be incorporated in GrEv to further improve performance.

\section{Conclusion}
\label{sec:conclusion}

We describe GrEv, a novel compiler for gradually-typed languages that focuses on structural, fine-grained, sound gradual typing. GrEv targets a low-level, unsafe backend (LLVM), and therefore the gradual typing machinery is responsible for performing type-related safety checks. What distinguishes GrEv from Grift is that it embraces the evidence-based semantics that are derived by the general framework of Abstracting Gradual Typing (AGT). The performance evaluation of GrEv on the Grift benchmarks shows that it is indeed possible to use evidence to compile gradual programs in a manner that is generally competitive, and can often be more efficient, than implementations based on coercions. Overall GrEv also exhibits much better stability across the typing precision spectrum. This is likely a consequence of the simpler and more uniform representation of evidence values in GrEv, compared to the coexistence of different representations in Grift.

During the design and implementation of GrEv we have also identified two novel semantic ideas related to monotonic references: first, we observe that performing in-place updates of evidence, in a shallow manner, which avoids the complexity of deep updates, retaining only the possibility to avoid proxies but not to avoid all runtime checks, is a reasonable alternative. Second, we have proposed monotonic closures, as the natural extension of the idea of monotonic references to first-class functions, which are also heap-allocated values. Monotonic designs, both in GrEv and Grift, can exhibit better performance than using proxies, although the gains we observe are neither universal nor always noticeable. This suggests that there is space for further exploration of this language design space, driven by a wider range of performance evaluations.

Likewise, our experience with the Dual Float Optimization (DFO) suggests that, while it does achieve top performance on fully-typed configurations, the performance degradation on partially-typed programs makes it less predictable compared to simply using heap-allocated floats. This confirms that there is a tension between aiming at full performance on statically-typed programs and average performance on the whole range of partially-typed programs. 
Here again, future work is needed to better understand how best to balance this tension.

In addition to the above, and incorporating several pending optimizations from the literature, this work opens up many venues for future work. Most notably, the current implementation of GrEv only supports simple gradual types with consistency. We believe that GrEv constitutes an appealing general infrastructure for studying the low-level implementations of a wide range of gradual typing disciplines. Given that evidence representation and consistent transitivity implementations are treated abstractly by a large part of the compilation pipeline, we can exploit the numerous applications of AGT in different settings, be it consistent subtyping, gradual effects, gradual parametricity or gradual sensitivity typing, among others.


\bibliography{extracted.bib}

\end{document}